\documentclass{article}

\usepackage{arxiv}

\usepackage[utf8]{inputenc} 
\usepackage[T1]{fontenc}    
\usepackage{hyperref}       
\usepackage{url}            
\usepackage{booktabs}       
\usepackage{amsfonts}       
\usepackage{nicefrac}       
\usepackage{microtype}      
\usepackage{lipsum}
\usepackage{times}
\usepackage{epsfig}
\usepackage{graphicx}
\usepackage{amsmath}
\usepackage{amsthm}
\usepackage{amssymb}
\usepackage{bm}
\usepackage{nccmath}
\usepackage{multirow}
\usepackage{url}
\usepackage{microtype}
\usepackage{algorithm2e}

\def \r{\mathbf{r}}

\newcommand{\genm}[3]{
	\left[\begin{array}{cc}
		#1  &  #3 \\
		#3 & #2 
	\end{array}
	\right]
}

\newcommand{\vecthe}[3]{
	\left[
	\begin{array}{c}
		#1 \\
		#2 \\
		#3
	\end{array}
	\right]
}

\newtheorem{prop}{\bf Proposition}
\graphicspath{{Figures/}}

\title{Image Restoration by  Combined Order Regularization with Optimal Spatial Adaptation}

\author{
  	Sanjay~Viswanath\\
  	Imaging Systems Lab\\
  	Department of Electrical Engineering\\
  	Indian Institute of Science\\
  	Bangalore, Karnataka, India 560012 \\
  	\texttt{sanjayv@iisc.ac.in} \\
  	\And
  	Manu~Ghulyani\\
  	Imaging Systems Lab\\
  	Department of Electrical Engineering\\
  	Indian Institute of Science\\
  	Bangalore, Karnataka, India 560012 \\
  	\texttt{sanjayv@iisc.ac.in} \\
  	\And
  	Simon~De~Beco\\
  	Laboratoire Physico Chimie Curie\\
  	Institut Curie\\
  	26 rue d’Ulm\\
  	Paris Cedex 05, France 75248\\
   	\texttt{simondebeco@gmail.com} \\
  	\And
  	Maxime~Dahan*\\
  	Laboratoire Physico Chimie Curie\\
  	Institut Curie\\
  	26 rue d’Ulm\\
  	Paris Cedex 05, France 75248\\
  	\And
	Muthuvel~Arigovindan\\
	Imaging Systems Lab\\
	Department of Electrical Engineering\\
	Indian Institute of Science\\
	Bangalore, Karnataka, India 560012 \\
	\texttt{mvel@iisc.ac.in} \\
	\And
	{\em *Dedicated to the memory of Maxime Dahan who passed away on 
			29.07.2018
	\centering}
}

\begin{document}
\maketitle

\begin{abstract}
	Total Variation (TV) and related extensions have been popular in image 
	restoration due to their robust performance and wide applicability. While
	the original formulation is still relevant after two decades of extensive 
	research, its extensions that combine derivatives of first and 
	second orders are now being explored for better performance, with examples
	being Combined Order TV (COTV) and Total Generalized Variation 
	(TGV).  As an improvement over such multi-order convex formulations, 
	we propose a novel non-convex regularization functional which 
	adaptively combines Hessian-Schatten (HS) norm and first order TV (TV1) 
	functionals with spatially varying weight. This  adaptive weight   
	itself is controlled by another regularization term;  the total
	cost becomes the sum of this adaptively weighted HS-TV1 term, 
	the regularization term for the adaptive weight, and the data-fitting
	term.  The reconstruction is obtained by jointly minimizing w.r.t. 
	the required image and the adaptive weight.  We construct a block 
	coordinate descent method for this minimization with proof of convergence,
	which alternates between minimization w.r.t. the required image and 
	the adaptive weights.  We derive exact computational formula for 
	minimization
	w.r.t.  the adaptive weight,  and construct an ADMM algorithm for 
	minimization
	w.r.t. to the required image.   We compare the proposed   
	method with existing    regularization methods,  and a recently proposed
	Deep GAN method using   image recovery   examples including   MRI   
	reconstruction and 
	microscopy   deconvolution.
\end{abstract}


\footnotetext{
	© 2020 IEEE.  Personal use of this material is permitted.  
	Permission from IEEE must be obtained for all other uses, in any current or 
	future media, including reprinting/republishing this material for 
	advertising 
	or promotional purposes, creating new collective works, for resale or 
	redistribution to servers or lists, or reuse of any copyrighted component 
	of 
	this work in other works.}

\section{Introduction}

Regularization  plays an important role in image
reconstruction/restoration by stabilizing the inversion of the imaging
forward model against noise and other distortions in modalities such as
photography \cite{CS_Baraniuk}, microscopy \cite{Microscopy_Mvel}, 
astronomical imaging \cite{astronomy_restoration} and 
magnetic resonance imaging 
\cite{Lustig_SparseMRI}.   Here we consider the measurement  model in
which  the measured image is expressed as a convolution of the underlying
image with a blurring function.  In regularized reconstruction,  the required 
image is the solution of the
following minimization problem:
\begin{equation}
	\label{eq:regrec}
	\hat{s}(\r) = \arg \min_{s} F(s,h,m) + \lambda R(s),
\end{equation}
where $R(s)$ is the regularization functional,  and $F(s,h,m)$ is the 
data fidelity term with $m$ being the measured image and $h$ being the
blurring kernel of the imaging system. We restrict the data fidelity term to be 
of the form,
\begin{equation}
	\label{eq:datafit}
	F(s,h,m) = \sum_{\bf r}|(h*s)({\bf r}) - m({\bf r})|^2
\end{equation}
Strictly speaking, the data-fidelity term should be the negative
log-likelihood  of the noise model.  However,  our focus is
on developing improved regularization, and hence we  use the above form
of data-fidelity as an approximation, irrespective of the noise model.  In our 
evaluation
of the regularized image reconstruction methods,  we consider two types of 
imaging
forward models with respect to our restoration experiments.  The first 
measurement model is related to total internal reflection fluorescence (TIRF) 
microscopy and is given by
\begin{equation}
	m(\r) = {\mathcal P}((h\ast s)(\r))  +  \eta(\r),
\end{equation} 
where $h(\r)$ is the PSF of TIRF microscope,   ${\mathcal P}(\cdot)$ represents 
the Poisson process, and  $\eta(\r)$
is additive white Gaussian noise.  The second measurement model involved in our 
experiments 
corresponds to   magnetic resonance imaging (MRI) and is given by 
\begin{equation}
	\hat{\bf m} = {\cal T}_s M{\cal F}s(\r)  +  \hat{\pmb \eta},
\end{equation} 
where  ${\cal F}$ represents Fourier transformation operation,  $M$ represents
the under-sampling mask of ones and zeros, 
${\cal T}_s$  represents the operations of picking samples from non-zero 
locations
of $M$,   and $\hat{\pmb \eta}$  is the complex  noise 
vector where each component comes from a Gaussian distribution.  For the first 
case,  equation (\ref{eq:datafit})  is not the exact negative log-likelihood, 
and 
for the second case, it is the exact negative log-likelihood.  For the MRI 
model in the second case, 
$h$ is the inverse Fourier transform of $M$  and $m(\r) =
{\cal F}^{-1}({\cal T}^{\dagger}_s\hat{\bf m})$,
with ${\cal T}^{\dagger}_s$ denoting the adjoint of ${\cal T}_s$,  which is 
essentially
the operation of embedding the Fourier samples into  image form.  Note that,
in this case,  $m(\r)$ can have complex values and it is taken care by the 
operator $|\cdot|$  in equation \eqref{eq:datafit}.

As mentioned before,  our focus is on developing an improved regularization 
method. 
The quality of restoration is mainly determined by the ability of $R(s)$ to 
discriminate between characteristics of the underlying image and noise. 
While priors can be defined based on general image characteristics such as 
sparsity of the roughness 
\cite{Vogel_TV_98,TV_App_BlindDecon, 
	TV_App_Wavelet, TV_App_Wlinpaint},
there are also priors which are tailored to specific classes of images 
\cite{k-svd, MRI_Bresler, Manifold_prior}.
The latter type of regularization approach    utilizes learning paradigms 
including deep learning based techniques \cite{DAGAN, Deep_MRI_1, Deep_MRI_2, 
Deep_MRI_3},  dictionary learning 
\cite{k-svd, MRI_Bresler, DL_Caballero14, DL_mubarakhigher} and  model fitting 
\cite{Manifold_prior}.  Here the functionals are built from training
images and then applied for restoration involving images from the same class. 
Such methods are believed  to 
out-perform general priors, when suitable training sets are available. At the 
same time, the necessity of 
training set and computational complexity limits their applicability.  

On the other hand, general priors such 
as Tikhonov regularization \cite{tikh_1} and total variation (TV) 
\cite{Rudin_tv1} do not need training 
samples and have been applied with robust performance in multiple domains. 
Among these general priors, 
total variation (TV) \cite{Rudin_tv1} has been widely applied 
\cite{Vogel_TV_98,TV_App_BlindDecon, 
	TV_App_Wavelet, TV_App_Wlinpaint} because of  its ability to recover sharp 
	image features in the
presence of noise and in the cases of undersampling. First order TV (TV1) 
restoration is given by 
\begin{equation}
	\begin{gathered}
		\label{eq:tv1}
		s_{opt} = \underset{s}{\operatorname{argmin}}\;\;F(s,h,m) + \lambda
		\underbrace{ \sum_{\bf r} \|({\bf d}_1*s)({\bf r}) \|_2 }_{R_1(s)} \\
		\mbox{with}\;\; d_{1}(\r) = [d_{x}(\r), d_{y}(\r)]^{T},
	\end{gathered}
\end{equation}
where $d_{x}(\r)$ and $d_{y}(\r)$ are filters implementing first order 
derivatives,
$\frac{\partial}{\partial x}$, and $\frac{\partial}{\partial y}$ respectively. 
While TV1 is able to retain edges \cite{TV_Edge} in the reconstruction as 
compared to standard $\ell_{2}$ norm based Tikhonov regularization 
\cite{tikh_1}, it presents drawbacks such as
staircase artifacts \cite{TV_Ring_Stair, TV_Stair_Jalalzai}.   
Higher order extensions of TV \cite{TV_Chambolle_97, Scherzer_tv2_98, 
	TV_Chan_2000, HS_13} have been proposed to avoid staircase artifacts and 
	they
deliver better restoration, albeit at the cost of increased computations. 
Second order TV (TV2) \cite{Scherzer_tv2_98} restoration was proposed as 
\begin{equation}
	\begin{gathered}
		\label{eq:tv2}
		s_{opt} = \underset{s}{\operatorname{argmin}}\;\;F(s,h,m) + \lambda
		\underbrace{\sum_{\bf r}\|({\bf d}_2*s)({\bf r})\|_2}_{R_2(s)}, \\
		\mbox{with}\;\; {\bf d}_2({\bf r}) = [d_{xx}({\bf r})\;\;d_{yy}({\bf 
		r})\;\;\sqrt{2}d_{xy}({\bf r})]^T,
	\end{gathered}
\end{equation}
where $d_{xx}(\r), d_{yy}(\r)$, and $d_{xy}(\r)$ are discrete filters 
implementing second order derivatives 
$\frac{\partial^2}{\partial x^2}$, $\frac{\partial^2}{\partial y^2}$ and 
$\frac{\partial^2}{\partial x\partial y}$ 
respectively. Another second-order derivative based formulation is 
Hessian-Schatten (HS) norm regularization \cite{HS_12} \cite{HS_13}, which has 
been proposed as a generalization
of the standard TV2 regularization.  It is constructed as an $\ell_p$ norm of 
Eigen values of the Hessian matrix, which
becomes the standard TV2 for $p=2$.  HS norm with $p=1$ has been proven to 
yield best resolution
in the reconstruction, since this better preserves Eigen values   of the 
Hessian \cite{HS_13}.
Let  ${\bf H}_2({\bf r})$ be  the matrix filter  composed of   $d_{xx}({\bf 
r})$, $d_{yy}({\bf r})$, 
and $d_{xy}({\bf r})$  and let ${\pmb \zeta}(\cdot)$ be the operator that 
returns the vector containing
the Eigen values of its matrix argument. Then HS norm regularization of order 
$p$ can be expressed as
\begin{equation}
	\label{eq:hs_reg}
	s_{opt} = \underset{s}{\operatorname{argmin}}\;\;F(s,h,m) + \lambda
	\sum_{\bf r}\left\|{\pmb \zeta}(({\bf H}_2*s)({\bf r}))\right\|_p.
\end{equation}
Since the Eigen values are actually directional second derivatives taken along 
principle directions, setting $p=1$
better preserves the local image structure. It has to be noted that the costs 
given in the equations  (\ref{eq:tv1}), and (\ref{eq:tv2})   are often 
minimized using 
gradient based approaches with smooth approximations of the form $R_k(s) = 
{\sum \limits_{\r}} \sqrt{\epsilon + 
	||d_{k}(\r) \ast s(\r)||_{2}^2}, \ k=1,2$ where $\epsilon$ is a small 
	positive constant  \cite{Rudin_tv1}, \cite{vogel_tv1}.
This approach  has been proven to converge to the minimum of the exact form as  
$\epsilon\rightarrow 0$
\cite{vogel_tv1}.  Approaches to minimize the cost without smooth approximation 
include primal-dual
method \cite{TV_Dual_Chambolle},  and  alternating direction of multiplier 
method (ADMM)  \cite{tvadmm}.
A detailed comparison of such approaches  has been provided in \cite{TV_Aujol}. 

It has been demonstrated that combining first- and second-order
derivatives is advantageous in accurately restoring image features 
\cite{TV_CombinedLysaker, Combined_order_TV, TGV, sam-tv}. 
In this regard, the combined order TV \cite{Combined_order_TV} uses scalar 
relative weights for combining first- and second-order variations, with the 
relative weights left as user parameters and the solution is estimated by means 
of optimization problem of
the form 
\begin{equation}
	s_{opt} = \underset{s}{\operatorname{argmin}}\;\;F(s,h,m)
	+ \lambda \alpha_1 R_1(s) + \lambda \alpha_2 R_2(s),
\end{equation} 
where $\alpha_1$  and $\alpha_2$ determine the relative weights.
Although HS norm can be in principle combined with TV1 by the same
way  as standard TV2 is combined  
as given above,  this possibility has not been 
explored.  A generalization for total variation to higher order terms, named as 
total generalized variation (TGV)  has also been proposed \cite{TGV} 
\cite{TGV2}.  It is generalized in the following ways:  it is formulated for any
general derivative order,  and for any given order,  it is generalized in the 
way
how the derivatives are penalized.  In the literature, only the second order 
TGV form has been
well explored for image reconstruction, which takes the following form: 
\begin{equation}
	\begin{split}
		(s_{opt}, {\bf p}_{opt}) & = \underset{s, {\bf 
		p}}{\operatorname{argmin}}\;\;F(s,h,m)  \\&+ 
		\lambda
		\alpha_1\sum_{\bf r} \|({\bf d}_1*s)({\bf r})-{\bf p}({\bf r})\|_2 \\
		&+ \lambda\alpha_2 
		\frac{1}{2} \sum_{\bf r}  \|{\bf d}_1({\bf r})*{\bf p}^T({\bf r})+{\bf 
		p}({\bf r})*{\bf d}_1^T({\bf r})\|_F,
	\end{split}
\end{equation}
where ${\bf p}({\bf r})$ is an auxiliary $2\times 1$ vector image.
The TGV functional is able to spatially adapt to the underlying image structure 
because of the minimization w.r.t. auxiliary variable
${\bf p}$. Near edges, ${\bf p}({\bf r})$ approaches zero leading to TV1-like 
behavior which allows sharp jumps in
the edges.  On the other hand,  in smooth regions,  ${\bf p}({\bf r})$ 
approaches  ${\bf d}_1*s({\bf r})$ leading
to TV2-like behavior which will avoid staircase artifacts. However, the 
drawback with TGV
functional is that   the weights  $\alpha_1$  and $\alpha_2$ have to be chosen 
by the user.
In summary, existing combined order methods can bring the advantages of 
multi-order derivatives albeit at the cost of additional tuning.

We propose a novel spatially adaptive regularization method, in which,  the  
weights involved in combining first- and second-order derivatives are determined
from the measured image without user-intervention.  Here the  relative weight 
between first- and second-order  terms 
becomes an image, and this weight is determined without user-intervention 
through minimization of a
composite cost function.  Our contributions can be summarized as follows:
\begin{itemize}
	\item
	We construct a composite regularization functional containing  two parts: 
	(i) the first part is constructed
	as the sum  of first- and second-order  derivative magnitudes with 
	spatially varying relative weights;
	(ii)  the second part is an additional regularization term for preventing 
	rapid spurious variations in the relative weights.  
	For the first order term, we use 
	the norm of the gradient, and for the second order term, we use the 
	Schatten norm of the Hessian 
	\cite{HS_12} \cite{HS_13}.
	The composite cost  functional is convex with respect to either the 
	required image or the relative weight,
	but it is non-convex jointly. 
	\item
	We construct a block coordinate descent method involving minimizations 
	w.r.t.  the required image and
	the relative weight alternatively  with the following structure: the 
	minimization w.r.t.  the required image
	is carried out using  ADMM approach \cite{eckstein1992douglas, 
	eckstein2017approximate}
	and the minimization w.r.t. the relative weight is carried out as a single 
	step exact minimization
	using a formula that we derive in this  paper. 
	\item
	Since the total cost is non-convex,  the reconstruction results are highly 
	dependent on the initialization
	for block-coordinate descent method.  We handle this problem using 
	a multi-resolution approach,  where, a series of coarse-to-fine 
	reconstructions are performed
	by minimization   of cost functionals defined through upsampling 
	operators.  Here, minimization w.r.t.  the relative weight and the  
	required image is carried out alternatively, as we progress
	from coarse to final resolution levels.  At the final resolution level, the 
	above-mentioned 
	block coordinate descent method is applied. 
	\item
	Note that  the sub-problem  of  minimization w.r.t. to the required image   
	involves spatially varying
	relative weights. Further, this sub-minimization problem  in the 
	above-mentioned multi-resolution
	loop involves upsampling operators.   Hence, the ADMM method proposed by 
	Papafitsoros et al.
	\cite{Combined_order_TV}    turns out to be unsuitable. We propose
	improved variable splitting method and computational formulas to handle 
	this issue.
	\item
	We prove that the overall block coordinate descent method converges to a 
	local minimum of the total
	cost function by using Zangwill's convergence theorem. 
\end{itemize}

This work is an extension of the work presented in the conference paper  
\cite{sam-tv}, where we only
considered a differentiable approximation of the total variation functionals, 
and did not incorporate
the joint optimization of the relative weight and the required image.  In other 
words, only the
multi-resolution loop is applied without the block-coordinate descent method in 
the final resolution.
This method was named Spatially Adaptive Multi-order TV (SAM-TV). Here we name 
the improved approach
more appropriately as  Combined Order  Regularization with Optimal Spatial 
Adaptation  (COROSA). 
The rest of the paper is organized as follows: section \ref{sec:corosa} deals 
with the formulation of COROSA functional and the multi-resolution framework, 
and section \ref{sec:admm} presents the ADMM formulation and iterations for the 
optimization problems associated with COROSA.  Section \ref{sec:experiments} 
presents the simulation results and comparisons.

\section{COROSA Image Restoration}
\label{sec:corosa}
\subsection{COROSA formulation}
In the proposed  COROSA  approach,  the restoration problem is formulated
as given below:
\begin{equation}
	\label{eq:corosa_cost_both}
	\begin{split}
		(s_{opt},\beta_{opt}) & = \underset{s,\beta}
		{\operatorname{argmin}} \;\; F(s,h,m) +  
		\lambda R_{sa}(s,\beta, p)  +  L(\beta, \tau) + {\cal B}(s),
	\end{split}
\end{equation}
where 
\begin{equation}
	\label{eq:rec}
	\begin{split}
		R_{sa}(s,\beta,  p) & = \sum_{\bf r}\beta({\bf r}) \|({\bf d}_1*s)({\bf 
		r}) \|_2  
		+ \sum_{\bf r}
		(1-\beta({\bf r}))\left\|{\pmb \zeta}(({\bf H}_{2}*s)({\bf 
		r}))\right\|_p,   \\
		&  \mbox{subject to} \;\; 0\le \beta(\mathbf{r}) \le 1,
	\end{split}
\end{equation}
$L(\beta, \tau)$ is a regularization term for $\beta$, which
will be specified soon, 
and ${\cal B}(\cdot)$ is the indicator function for constraining the 
restored image to a particular range of positive values.  From this
formulation,   it is clear that  the relative weight image $\beta$ 
is also considered as a minimization variable  and the optimal 
image of weights is determined jointly with the required image.
Since  $R_{sa}(s,\beta,  p)$ is linear in $\beta$,   
minimizing with respect to $\beta$  means that it will essentially
act as a switching  between first- and second-order terms.
In this context, the role of $L(\beta, \tau)$ is to  prevent spurious
switching;  in other words, its role is to prevent rapid switching 
between the first- and second-order terms  caused by insignificant
differences in their magnitudes.  We set
$L(\beta, \tau)$  as 
\begin{equation}
	\label{eq:lbdef}
	L(\beta, \tau) = - \sum_{\bf r} 
	\tau\log(\beta({\bf r})(1-\beta({\bf r})))	.
\end{equation}
Here,  a lower  value of $\tau$ will cause a more rapid
switching  between first- and second-order terms and vice versa.
We denote the overall
cost by $J_{sa}(s, \beta, \tau, h, m)$,  and   we write
\begin{equation}
	\label{eq:fullcorosacost}
		J_{sa}(s, \beta, \tau, h, m) = 
		F(s,h,m) +  
		\lambda R_{sa}(s,\beta, p) 
		+  L(\beta, \tau) + {\cal B}(s).
\end{equation}
An assumption that is implicitly made by most of the image restoration 
algorithms
is that there is no $s$ for which $F(s,h,m)$, 
$ R_1(s) = \sum_{\bf r}  \|({\bf d}_1*s)({\bf r}) \|_2$, and 
$R_2(s) = 
\sum_{\bf r}  \|({\bf H}_2*s)({\bf r}) \|_2$ will have zero value 
simultaneously.
We will also use this assumption  for proving the convergence of the iterative 
method
that we propose in the following sections.


\subsection{Multiresolution  method}
\label{sec:mrm}

The regularization functional, $R_{sa}(s,\beta,  p)$,    is non-convex
jointly with respect to $\beta$ and $s$, although it is convex with any one of
them alone.    Hence, the reconstruction result becomes sensitive to 
initialization and finding an efficient initialization becomes crucial.
To this end, we adopt multiresolution approach for the initialization. To 
describe the 
multi-resolution approach,  we define the following:
\begin{equation}
	F^{(j)}(s, h, m) = \sum_{\bf r}|(h*(E^{(j)}s))({\bf r}) - m({\bf r})|^2.
\end{equation}
\begin{equation}
	\label{eq:rsamr}
	\begin{split}
		R^{(j)}_{sa}(s,\beta,  p) & = \sum_{\bf r}\beta({\bf r}) 
		\|({\bf d}_1*(E^{(j)}s))({\bf r}) \|_2  
		\sum_{\bf r}
		(1-\beta({\bf r}))\left\|{\pmb \zeta}
		(({\bf H}_{2}*(E^{(j)}s))({\bf r}))\right\|_p.
	\end{split}
\end{equation}
In the above,  $E^{(j)}s$   denotes the image obtained by interpolating
$s$  by a factor $2^j$ along both axes.    We will defer the description
of the implementation of $E^{(j)}$ to the end. 
Here,  $(h*(E^{(j)}s))({\bf r})$
denotes convolving the interpolated image $E^{(j)}s$   with $h$ 
followed by accessing the pixel at position $\mathbf{r}$. Further, 
$({\bf d}_1*(E^{(j)}s))({\bf r})$  and
$({\bf H}_{2}*(E^{(j)}s))({\bf r}))$
have similar interpretation except that the first expression
will be  a vector,  and the second one will be a matrix.  
Note that the variable $s$ is an $\frac{N}{2^j}\times \frac{N}{2^j}$ image
in both the functionals  $F^{(j)}(s, h, m)$, and $R^{(j)}_{sa}(s,\beta, \tau, 
p)$.
On the other hand, $\beta$ always has size
${N}\times {N}$, which is the size of the measurement
$m(\mathbf{r})$.  We denote the resulting scale-$j$  cost by
\begin{equation}
	\label{eq:fullcorosacostj}
	\begin{split}
		J_{sa}^{(j)}(s, \beta, \tau, h, m) & = 
		F^{(j)}(s,h,m) +  
		\lambda R_{sa}^{(j)}(s,\beta, p)  
		+  L(\beta, \tau) + {\cal B}(E^{(j)}s).
	\end{split}
\end{equation}
It should be noted that,  size of the variable in scale-$j$  cost is 
$\frac{N}{2^j}\times \frac{N}{2^j}$; however, the cost is always
evaluated on a ${N}\times {N}$ grid through interpolation
by $E^{(j)}$.  This will help to ensure a better convergence in the
multi-resolution method to be described below.

Let $K$ denote the user-defined number of multi-resolution levels. 
To initialize the multi-resolution loop, we set
$\beta({\bf r})=0$ and perform the following
minimization:
\begin{equation}
	\label{eq:s_init}
	\begin{split}
		\hat{s}^{(K)}({\bf r}) & = \underset{s}
		{\operatorname{argmin}} \ J_{sa}^{(K)}(s, \beta, \tau, h, m) \\
		& \equiv  \underset{s}{\operatorname{argmin}} 
		\;\; F^{(K)}(s,h,m)  
		~~~~~~~~~~ +
		\lambda R^{(K)}_{sa}(s,\beta,  p) +  {\cal B}(E^{(K)}s).
	\end{split}
\end{equation}
With   $\hat{s}^{(K)}$ as the initialization, we iterate for $j=K-1,\ldots,0$ 
with the following minimizations:
\begin{align}
	\nonumber
	f & =E^{(j+1)}\hat{s}^{(j+1)}  \\
	\nonumber
	\bar{\beta}({\bf r}) & = \underset{\beta}{\operatorname{argmin}}  \;\;
	J_{sa}(f, \beta, \tau, h, m)  \\
	\label{eq:mrbeta}
	& \equiv \underset{\beta}{\operatorname{argmin}}  \;\;
	R_{sa}(f,\beta,  p)  +  L(\beta, \tau),   \\
	\nonumber
	&\; \mbox{subject to}\; 0\le \beta(\r) \le 1. \\
	\nonumber
	\hat{s}^{(j)}({\bf r}) & =  \underset{s}{\operatorname{argmin}} \;\;
	J_{sa}^{(j)}(s, \bar{\beta}, \tau, h, m)  \\
	\label{eq:mrsi}
	& \equiv \underset{s}{\operatorname{argmin}}
	\;\;F^{(j)}(s,h,m)   
	~~~~~ +  
	\lambda R^{(j)}_{sa}(s,\bar{\beta},  p) 
	+  {\cal B}(E^{(j)}s).
\end{align}
The resulting restored image at the end of the multi-resolution loop,
$\hat{s}^{(0)}$,  can be an initialization for the  joint minimization
problem given in equation \eqref{eq:corosa_cost_both}.
Note that,  the minimization problem of equation \eqref{eq:mrsi}
has to be done iteratively and requires an initialization.  We use
$E^{(1)}\hat{s}^{(j+1)}({\bf r})$ as the initialization.  In this regards,
the way the multi-scale costs  
$\{J_{sa}^{(j)}(s, \beta, \tau, h, m), j=0,\ldots,K\}$ are constructed
significantly helps to make the initialization $E^{(1)}\hat{s}^{(j+1)}({\bf r})$
to be very close the the required minimum $\hat{s}^{(j)}({\bf r})$.
In other words,  since each $J_{sa}^{(j)}$  is 
constructed through  $2^j$-fold interpolation  on the original reconstruction
grid,  the initialization $E^{(1)}\hat{s}^{(j+1)}({\bf r})$ is typically 
very close the the required minimum $\hat{s}^{(j)}({\bf r})$.

In the multi-resolution method described above,  if we set $p=2$,
the result $\hat{s}^{(0)}$  will be equivalent to final reconstruction
of  SAM-TV approach   \cite{sam-tv}  except the fact that 
SAM-TV uses smooth approximations for the first- and second-order 
TV terms,  whereas, here, the exact non-differentiable form of TV functionals
are used. Now we consider solving the sub-problem of determining the 
adaptive weight $\bar{\beta}$ in equation \eqref{eq:mrbeta}. The exact 
solution for $\bar{\beta}$ is  given in the following Proposition.      
\begin{prop}
	\label{prop:beta}
	Let $d(\r)$ be defined as
	\begin{equation}
		\label{eq:ddef}
		d(\r) = \|({\bf d}_1*f)({\bf r}) \|_2 - 
		\left\|{\pmb \zeta}(({\bf H}_{2}*f)({\bf r}))\right\|_p.
	\end{equation}
	If $d(\r) =0$, the solution for the problem in equation \eqref{eq:mrbeta} is
	$\bar{\beta}(\r) = 0.5$. When $d(\r)$ is non-zero,
	the solution $\bar{\beta}$  is unique and given by
	\begin{equation}
		\label{eq:betaexp}
		\bar{\beta}(\r) = \frac{1}{2} \left(1  - sign(d(\r))\left(  
		\sqrt{\frac{4\tau^{2}}{d^2(\r)} + 1}- 
		\frac{2\tau}{|d(\r)|}\right)\right).
	\end{equation}
\end{prop}
\noindent
The proof of Proposition \ref{prop:beta} is given in Appendix. 
Next, unlike equation \eqref{eq:mrbeta},  the subproblem of 
equation \eqref{eq:mrsi}  cannot be solved exactly and has to be solved
iteratively.   We will develop an ADMM based method to solve this problem
in Section \ref{sec:admm}.

Now it remains to specify the implementation of $E^{(j)}$.  It can be 
implemented by $j$ stages of $2$-fold interpolation.   The   $2$-fold 
interpolation can be implemented by inserting a zero next to each pixel 
along both axes (which is called as the two-fold expansion) and then filtering
by   an appropriate interpolation filter. In our implementation  we use
$u(\r) =  \frac{1}{64} [1\; 4 \; 6 \; 4 \; 1]^T[1\; 4 \; 6 \; 4 \; 1]$ as the 
interpolation
filter, which is the two-scale filter of cubic B-spline
\cite{unsersigpromag}.

\subsection{Obtaining the final restoration by block coordinate
	descent method}

By using the result of the above multiresolution method as the  initialization,
the final reconstruction has to be obtained by minimizing the cost of
equation \eqref{eq:corosa_cost_both} jointly with respect to 
$\beta$ and $s$.  We propose to use a simple block coordinate descent 
method.  Let $s_{(0)}=\hat{s}^{(0)}$, where  $\hat{s}^{(0)}$ is the result of
the multi-resolution loop described before. With $k=0,\ldots, N_b$,  the block 
coordinate descent method  involves the following series of   minimizations
with respect to $\beta$ and $s$.
Let $s_{(k)}$ and $\beta_{(k)}$
be the current estimate of the minimum at cycle $k$. Then 
the next refined estimate is computed as the following set of
minimizations:
\begin{align}
	\nonumber
	{\beta}_{(k+1)}  & = 
	\underset{\beta}{\operatorname{argmin}}\;\;
	J_{sa}(s_{(k)}, \beta, \tau, h, m)  \\
	\label{eq:bcd1}
	& = 
	\underset{\beta}{\operatorname{argmin}}\;\;
	R_{sa}(s_{(k)},\beta,  p) + L(\beta,\tau).
\end{align}
\begin{align}
	\nonumber
	s_{(k+1)} &= \underset{s}{\operatorname{argmin}} \;\;
	J_{sa}(s, \beta_{(k+1)}, \tau, h, m)  \\
	\label{eq:bcd2}
	&= 
		\underset{s}{\operatorname{argmin}} \; F(s,m,h) + \lambda
		R_{sa}(s,\beta_{(k+1)}, p) +  {\cal B}(s).
\end{align}
As evident,  the iterations given above are similar to the iterations
given in the multi-resolution method of section  \ref{sec:mrm}.
The difference is that the minimization with respect to $\beta$ and
$s$ for each of the cost functions in the series 
$\{J_{sa}^{(j)}(s, \beta, \tau, h, m), j=0,\ldots,K\}$ is done only once
in the multiresolution method.
On the other hand,  the minimizations in 
the block coordinate decent method (BCD) 
represented
by the equation \eqref{eq:bcd1} and \eqref{eq:bcd2}  are done alternatively
on the same cost function 
$J_{sa}(s, \beta, \tau, h, m)$ until convergence.
The functional $J_{sa}(s, \beta, \tau, h, m)$ is convex with respect to either 
of $s$ 
and $\beta$,  and  the  BCD method represented by equations
\eqref{eq:bcd1} and \eqref{eq:bcd2}   converges to the solution of the problem
given in \eqref{eq:corosa_cost_both},
provided that
each of the minimizations is exact  as per the convergence theorem of
Bertsekas \cite{Bertsekas_NLP}. 

Now we consider solving the sub-problems. The subproblems of determining the 
adaptive weight $\bar{\beta}$  in the block coordinate descent method 
of equation  \eqref{eq:bcd1}   is identical to the sub-problem of the 
multi-resolution
method  (equation  \eqref{eq:mrbeta}), and hence can be solved exactly.
On the other hand,  the sub-problem of equation \eqref{eq:bcd2} is similar to 
the
problem of   the equation \eqref{eq:mrsi},  and cannot be solved exactly.  
Hence the convergence result of Bertsekas \cite{Bertsekas_NLP} will not  be 
applicable. 
However,  it is easy to show that BCD iteration converges to the minimum 
if $J_{sa}(s_{(k+1)}, \beta_{(k+1)}, \tau, h, m)  < 
J_{sa}(s_{(k)}, \beta_{(k+1)}, \tau, h, m)$ using Zangwill's global convergence
theorem.   We will provide the convergence statement along with the proof
after describing the ADMM method for solving the problems of equations
\eqref{eq:mrsi}  and \eqref{eq:bcd2} in the next section.

\section{Image recovery with fixed relative weight for  first- and
	second-order derivatives }
\label{sec:admm}

The main computational  task in the block coordinate descent
iteration represented by the equations \eqref{eq:bcd1}  and 
\eqref{eq:bcd2}, is the computation of $s_{(k+1)}$.
Similarly,  in the multi-resolution method represented by equations
\eqref{eq:mrbeta} and \eqref{eq:mrsi},  the main task
is the computation of $\hat{s}^{(j)}({\bf r})$.
Note that the cost in    \eqref{eq:mrsi}  becomes algebraically 
identical to the cost of \eqref{eq:bcd2}    for $j=0$.
Hence the cost in equation \eqref{eq:bcd2}  can be considered
as a special case of the cost in the equation \eqref{eq:mrsi}.
So we consider only  the   description of the minimization of the
cost in \eqref{eq:mrsi}.  
We will use the ADMM approach for solving the minimization 
problem given in the equation \eqref{eq:mrsi}.  Here, the result of
previous level $j+1$, denoted by $\hat{s}^{(j+1)}$,  can be used
for initializing after interpolating  by factor of two.    
Although ADMM
is well-known and its application for total variation based 
image restoration is not new \cite{tvadmm, TV_ADMM_Afonso, 
	TV_ADMM_Steidl, TV_ADMM_Manu},
implementation of standard ADMM causes some numerical 
problems because of the spatially varying relative weight $\bar{\beta}$.
In the following, we will first describe the formulation that will lead to
standard  ADMM and then describe the modification necessary to handle 
the associated  numerical issues.  The first step  in constructing an ADMM 
algorithm for 
minimizing composite functionals is to define an equivalent constrained 
optimization
such that the  sub-functionals  act on different set of variables that are 
related
by means of linear equality constraints.  Then writing the augmented Lagrangian 
\cite{Bertsekas_NLP}  for the constrained problems leads to the required ADMM 
algorithm.


\subsection{Constrained formulation and variable splitting}

For notational convenience, we switch to vector based notations. 
Let the $N\times N$ image $s({\bf r})$ be represented by scanned 
vector ${\bf s}$ in $\mathbb{R}^{N^2}$, such that its $i$th element 
$s_i$ is given by $s_{i({\bf r}^{\prime})}=s({\bf r}^{\prime})$ with  ${\bf 
r}^{\prime}=
[r_1\; r_2]$  satisfying $i({\bf r}^{\prime})=r_2N+r_1$. Let ${\bf m}$ and
$\bar{\pmb \beta}$ also be  defined from $m({\bf r})$ and 
$\bar{\beta}(\r)$ in a similar way with the components denoted by $m_i$ and 
$\bar{\beta}_i$.
Let ${\bf H}$ be the matrix 
equivalent of convolving an image with $h({\bf r})$, such that  the 
scanned vector of $(h*s)({\bf r})$ is given by ${\bf H}{\bf s}$. Next, let
${\bf E}^{(j)}$ be   the matrix equivalent of interpolation by a 
factor $2^i$. With this, the scanned vector
of $E^{(j)}s({\bf r})$ is given by ${\bf E}^{(j)}{\bf s}$.  In terms of the new
notational scheme, the  data 
fidelity term can be   written as
\begin{equation}
	F^{(j)}(\mathbf{s,H,m})  = \left\|{\bf HE^{(j)}s} - {\bf m}\right\|^2_2.
\end{equation}
Similarly, let ${\bf D}_x$, ${\bf D}_y$, ${\bf D}_{xx}$, ${\bf D}_{yy}$, and 
${\bf D}_{xy}$ be the matrices defined from $d_x({\bf r})$, $d_y({\bf r})$, 
$d_{xx}({\bf r})$, $d_{yy}({\bf r})$, and $d_{xy}({\bf r})$ for representing 
convolution operations. Let ${\bf D}_f = 
[{\bf D}_x^T \; {\bf D}_y^T]^T$ and let ${\bf D}_s = [{\bf D}_{xx}^T \;  
{\bf D}_{yy}^T \; {\bf D}_{xy}^T]^T$.
Let  ${\cal S}({\bf v}): \mathbb{R}^3\rightarrow \mathbb{R}^4$ be the 
mapping that returns $\genm{v_1}{v_2}{v_3}$, where ${\bf v} = 
\{v_1,v_2,v_3\} \in \mathbb{R}^3$ represents the three second order 
derivatives.
Let ${\bf P}_i$ be a $2\times 2N^2$ matrix having ones at  locations 
$(1,i)$, and $(2,N^2+i)$ and zeros at all other locations.   
Let ${\bf Q}_i$ be the $3\times 3N^2$ matrix having ones at  locations 
$(1,i)$, 
$(2,N^2+i)$, and $(3,2N^2+i)$
and zero at other locations. 
Then we can use the following substitutions  in Eq.\eqref{eq:rsamr}:
\begin{equation}
	\|({\bf d}_1*(E^{(j)}s))({\bf r})\|_2 = \left\|{\bf P}_{i({\bf r})}{\bf 
	D}_f\mathbf{E}^{(j)}{\bf s}\right\|_2 
\end{equation}
\begin{equation}
	\left\|{\pmb \zeta}(({\bf H}_2*(E^{(j)}s))({\bf r}))\right\|_p= 
	\left\|{\pmb \zeta}({\cal S}({\bf Q}_{i({\bf r})}{\bf 
	D}_s\mathbf{E}^{(j)}{\bf s}))\right\|_p 
\end{equation}
With these,  the regularization functional can be expressed as
\begin{equation}
	\label{eq:recmat}
	R^{(j)}_{sa}({\bf s},{\pmb \beta}, p) =  \sum_{i=1}^{N^2}\bar{\beta}_i 
	\left\|{\bf P}_{i}{\bf D}_f\mathbf{E}^{(j)}{\bf s}\right\|_2 
	+ \sum_{i=1}^{N^2}(1-\bar{\beta}_i) \left\|{\pmb \zeta}({\cal S}({\bf Q}_i
	{\bf D}_s\mathbf{E}^{(j)}{\bf s}))\right\|_p,
\end{equation}
where we have replaced the images $s$ and $\bar{\beta}$ by 
the components of 
their vectorial form.  
Next,  to simplify the task of expressing and comparing  
the two forms of ADMM,   we introduce two more definitions as given below:
\begin{align}
	\mathcal{N}_f(\mathbf{u},\mathbf{v}) 
	& = \sum_{i=1}^{N^2} u_i \left\|{\bf P}_{i} \mathbf{v}\right\|_2  \\
	\mathcal{N}_s(\mathbf{u},\mathbf{w},p)
	& =   \sum_{i=1}^{N^2} u_i
	\left\|{\pmb \zeta}({\cal S}({\bf Q}_{i}\mathbf{w}))\right\|_p
\end{align}
Note that $\mathbf{u} \in \mathbb{R}^{N^2}$, 
$\mathbf{v} \in \mathbb{R}^{2N^2}$, and 
$\mathbf{w} \in \mathbb{R}^{3N^2}$.
With these, the regularization can be expressed  as
\begin{equation}
	\label{eq:regformn}
	\begin{split}
		\lambda R^{(j)}_{sa}({\bf s},{\pmb \beta}, p)
		= \mathcal{N}_f(\lambda {\pmb \beta},  
		\mathbf{D}_f\mathbf{E}^{(j)}\mathbf{s}) 
		+
		\mathcal{N}_s(\lambda ({{\bf 1} - {\pmb \beta}}),  
		\mathbf{D}_s\mathbf{E}^{(j)}\mathbf{s},p) 
	\end{split}
\end{equation}
where $\mathbf{1}$ is the vector of  ones.
Now the minimization of $F^{(j)}(\mathbf{s,H,m})
+  \lambda R^{(j)}_{sa}({\bf s},{\pmb \beta}, p) +  {\cal B}({\bf E}^{(j)}{\bf 
s})$ 
can be equivalently
expressed as
\begin{equation}
	\label{eq:admm_cost}
	\begin{split}
		({\bf s}^*, \mathbf{d}^*_f, \mathbf{d}^*_s,
		\mathbf{d}^*_0) 
		& = \underset{{\bf s}, \mathbf{d}_f, \mathbf{d}_s,
			\mathbf{d}_0}{\operatorname{argmin}} \;\;
		F^{(j)}(\mathbf{s,H,m}) + 
		\mathcal{N}_f(\lambda {\pmb \beta},  \mathbf{d}_f )  
		+
		\mathcal{N}_s(\lambda ({\bf 1}-{\pmb \beta}),  \mathbf{d}_s, p) 
		+{\cal B}({\bf d}_0),
	\end{split}
\end{equation}
subject to the conditions that 
\begin{equation}
	{\bf D}_f\mathbf{E}^{(j)}\mathbf{s}={\bf d}_f,	
	{\bf D}_s\mathbf{E}^{(j)}\mathbf{s}={\bf d}_s,
	\mathbf{E}^{(j)}\mathbf{s} = {\bf d}_0.
\end{equation}
This constrained formulation of the reconstruction problem leads to
the ADMM algorithm, which is essentially a series of cyclic minimization
of individual sub-functionals of the above cost. However,   through  some
reconstruction trials,  we found that ADMM  method  obtained from this
formulation leaves some 
artifacts in the reconstruction.  These artifact disappear only with very large 
number of 
iterations.  

Here we present an alternative formulation  that leads to a better 
converging ADMM algorithm.    To this end, we first recognize that
the cost given in the equation  \eqref{eq:regformn} can also be expressed as
\begin{equation}
	\label{eq:regformn2}
	\lambda R^{(j)}_{ec}({\bf s},{\pmb \beta}, p)
	= \mathcal{N}_f(\lambda  \mathbf{1},  
	\mathbf{D}_f^{\prime}\mathbf{E}^{(j)}\mathbf{s}) +
	\mathcal{N}_s(\lambda \mathbf{1},  
	\mathbf{D}_s^{\prime}\mathbf{E}^{(j)}\mathbf{s},p) 
\end{equation}
where
\begin{equation}
	\label{eq:dfp}
	{\bf D}^{\prime}_f = [{\bf D}_x^T{\bf B} \; {\bf D}_y^T{\bf B}]^T 
\end{equation}
\begin{equation}
	\label{eq:dsp}
	{\bf D}_s^{\prime} = [{\bf D}_{xx}^T({\bf I} - {\bf  B}) \;  
	{\bf D}_{yy}^T({\bf I} - {\bf  B}) \; {\bf D}_{xy}^T({\bf I} - {\bf  B})]^T
\end{equation}
with ${\bf B}=diag({\pmb \beta})$. 
The corresponding constrained problem becomes
\begin{equation}
	\label{eq:admm_cost2}
	\begin{split}
		({\bf s}^*, \mathbf{d}^*_f, \mathbf{d}^*_f,
		\mathbf{d}^*_0) 
		& = \underset{{\bf s}, \mathbf{d}_f, \mathbf{d}_f,
			\mathbf{d}_0}{\operatorname{argmin}} \;\;
		F^{(j)}(\mathbf{s,H,m}) + 
		\mathcal{N}_f(\lambda {\bf 1},  \mathbf{d}_f )  
		+
		\mathcal{N}_s(\lambda {\bf 1},  \mathbf{d}_s,p) 
		+{\cal B}({\bf d}_0),
	\end{split}
\end{equation}
subject to condition that 
\begin{equation}
	\label{eq:cns21}
	{\bf D}_f^{\prime}\mathbf{E}^{(j)}\mathbf{s}={\bf d}_f,
	{\bf D}_s^{\prime}\mathbf{E}^{(j)}\mathbf{s}={\bf d}_s,
	\mathbf{E}^{(j)}\mathbf{s} = {\bf d}_0.
\end{equation}
From our experiments, we found that the ADMM steps constructed based on the 
above constrained formulation
leads to better converging algorithm.  


\subsection{Augmented Lagrangian and the ADMM steps}

Writing the ADMM steps for the above
problem is straightforward and well-known in the literature. However,  for 
proving
the convergence of block coordinate descent method represented by equations
\eqref{eq:bcd1}  and \eqref{eq:bcd2},    we need to specify the steps here.
Further,  the constraint of the modified formulation given by equation
\eqref{eq:cns21}  involves non-circulant matrices and hence,
it requires some special consideration. 
To proceed further,  we use the symbol $\bar{\mathbf{s}}$ in the place of  
${\mathbf{s}}$ to avoid
notational conflict   with the iterations of equations 
\eqref{eq:mrbeta} and \eqref{eq:mrsi}.
To construct the ADMM algorithm from the above constraint form
of the problem, we
define
\begin{equation}
	\label{eq:cdef}
	\begin{split}
		C(\bar{\mathbf{s}}, \mathbf{d}_f, \mathbf{d}_s, \mathbf{d}_0,
		\mathbf{w})
		&  =
		\frac{\gamma}{2}
		\left(\left\| {\bf D}_f^{\prime}\mathbf{E}^{(j)}\bar{\mathbf{s}}
		-{\bf d}_f	\right\|_2^2  \right. 
		+
		\left\| {\bf D}_s^{\prime}\mathbf{E}^{(j)}\bar{\mathbf{s}}
		-{\bf d}_s	 \right\|_2^2   
		\left. +
		\left\| \mathbf{E}^{(j)}\bar{\mathbf{s}}
		-{\bf d}_0 \right\|_2^2 \right)  
		\\
		&  
		+ 
		\mathbf{w}^T\left( \vecthe{{\bf D}_f^{\prime}}{{\bf D}_s^{\prime}}{\bf 
		I}{\bf E}^{(j)}\bar{\mathbf{s}} -  
		\vecthe{{\bf d}_f}{{\bf d}_s}{{\bf d}_0} \right)
	\end{split}
\end{equation}	
where $\mathbf{w}$ represents   Lagrangian multipliers \cite{Bertsekas_NLP}. 
Let 
\begin{equation}
	\label{eq:jadef}
	\begin{split}
		\mathcal{J}_a(\bar{\mathbf{s}}, \mathbf{d}_f, \mathbf{d}_s,
		\mathbf{d}_0, \lambda)  & = 
		F^{(j)}(\mathbf{\bar{\mathbf{s}},H,m}) + 
		\mathcal{N}_f(\lambda {\pmb 1},  \mathbf{d}_f )  
		 +
		\mathcal{N}_s(\lambda {\pmb 1},  \mathbf{d}_s,p) 
		+{\cal B}({\bf d}_0),
	\end{split}
\end{equation}
and 
\begin{equation}
	\begin{split}
		L_a(\bar{\mathbf{s}}, \mathbf{d}_f, \mathbf{d}_s, \mathbf{d}_0,
		\mathbf{w},
		\lambda ) & =
		\mathcal{J}_a(\bar{\mathbf{s}}, \mathbf{d}_f, \mathbf{d}_f,
		\mathbf{d}_0, \lambda) 
		+ C(\bar{\mathbf{s}}, \mathbf{d}_f, \mathbf{d}_s, \mathbf{d}_0,
		\mathbf{w} )
	\end{split}
\end{equation}

Let $\bar{\mathbf{s}}^{(j)}$  be the vector of length $N^2/2^{2j}$  obtained
by scanning  $E^{(1)}\hat{s}^{(j+1)}(\mathbf{r})$, where
$\hat{s}^{(j+1)}(\mathbf{r})$ is the result of previous iteration in the
multi-resolution loop of the equations \eqref{eq:mrbeta}  and
\eqref{eq:mrsi}.
Then ADMM iterations
proceed as follows for $k=0,1,2,\ldots$:
\begin{align}
	\label{eq:dfmin}
	\mathbf{d}_f^{(k+1)} &  = \underset{{\bf d}_f}{\operatorname{argmin}}
	L_a(\bar{\mathbf{s}}^{(k)}, \mathbf{d}_f, \mathbf{d}_s^{(k)},
	\mathbf{d}_0^{(k)},
	\mathbf{w}^{(k)},
	\lambda)  \\
	\label{eq:dsmin}
	\mathbf{d}_s^{(k+1)} &  = \underset{{\bf d}_s}{\operatorname{argmin}}
	L_a(\bar{\mathbf{s}}^{(k)}, \mathbf{d}_f^{(k+1)}, \mathbf{d}_s,
	\mathbf{d}_0^{(k)},
	\mathbf{w}^{(k)},
	\lambda )  \\
	\label{eq:d0min}
	\mathbf{d}_0^{(k+1)} &  = \underset{{\bf d}_0}{\operatorname{argmin}}
	L_a(\bar{\mathbf{s}}^{(k)}, \mathbf{d}_f^{(k+1)}, \mathbf{d}_s^{(k+1)},
	\mathbf{d}_0,
	\mathbf{w}^{(k)},
	\lambda)  \\
	\label{eq:smin}
	\bar{\mathbf{s}}^{(k+1)} &  = \underset{\bar{\bf s}}{\operatorname{argmin}}
	L_a(\bar{\mathbf{s}}, \mathbf{d}_f^{(k+1)}, \mathbf{d}_s^{(k+1)}, 
	\mathbf{d}_0^{(k+1)},
	\mathbf{w}^{(k)},
	\lambda )  \\
	\label{eq:wmin}
	\mathbf{w}^{(k+1)} & = \mathbf{w}^{(k)} + \gamma 
	\left( \vecthe{{\bf D}_f^{\prime}}{{\bf D}_s^{\prime}}{\bf I}{\bf 
	E}^{(j)}\bar{\mathbf{s}}^{(k+1)} -  
	\vecthe{{\bf d}_f^{(k+1)}}{{\bf d}_s^{(k+1)}}{{\bf d}_0^{(k+1)}} \right) 
\end{align}
We will use the partitioned notation 
$\mathbf{w}^{(k)} = \vecthe{{\bf w}_f^{(k)}}{{\bf w}_s^{(k)}}{{\bf w}_0^{(k)}}$
for defining the solution for the above minimization
problems.

First note that 
$L_a(\bar{\mathbf{s}}, \mathbf{d}_f, \mathbf{d}_s, \mathbf{d}_0,
\mathbf{w}, \lambda )$ is non-differentiable w.r.t. $ \mathbf{d}_f$, 
$\mathbf{d}_s$,  and $\mathbf{d}_0$, and we use the notion of proximal
map to carry out the minimization defined in the equations
(\ref{eq:dfmin}),  (\ref{eq:dsmin}), and  (\ref{eq:d0min}).  
The minimization w.r.t.   $\mathbf{d}_f$  given in the equation 
(\ref{eq:dfmin}) is
essentially the minimization
of 
$\frac{\gamma}{2}
\left\| {\bf D}_f^{\prime}\mathbf{E}^{(j)}\bar{\mathbf{s}}
-{\bf d}_f	\right\|_2^2  +
(\mathbf{w}^{(k)}_f)^T ( {\bf D}_f^{\prime}{\bf E}^{(j)}\bar{\mathbf{s}} - {\bf 
d}_f) + 
\mathcal{N}_f(\lambda {\pmb 1},  \mathbf{d}_f )$,  w.r.t. ${\bf d}_f$.  This 
can be
rewritten as 
$\frac{\gamma}{2}
\left\|  \bar{\mathbf{d}}_f^{(k)}
-{\bf d}_f	\right\|_2^2  + 
\mathcal{N}_f(\lambda {\pmb 1},  \mathbf{d}_f ) + const. $, 
where 
$\bar{\mathbf{d}}_f^{(k)} = 
\mathbf{D}_f^{\prime}{\bf E}^{(j)}\bar{\mathbf{s}}^{(k)} 
+ (1/\gamma){\bf w}_f^{(k)}$, and   $const$ is a term that is independent of 
${\bf d}_f$.  The minimum of this cost is defined as the proximal map
of $\mathcal{N}_f(\lambda {\pmb 1},  \cdot )$ applied on 
$\bar{\mathbf{d}}_f^{(k)}$  \cite{proximal_parikh}.   This map is expressed as  
\cite{proximal_parikh}
\begin{equation}
	\label{eq:admm_df}
	\mathbf{d}_f^{(k+1)} = \sum_i^{N^2} \mathbf{P}_i^T
	{\cal T}(\mathbf{P}_i\bar{\mathbf{d}}_f^{(k)}, \lambda /\gamma),
\end{equation}
where $\mathcal{T}(\mathbf{x},t)$ denotes the soft-threshold operator
given by
\begin{equation}
	\label{eq:admm_T}
	\mathcal{T}(\mathbf{x},t)  = 
	max(0,\left\|\mathbf{x}\right\|_2-t)\frac{\mathbf{x}}
	{\left\|\mathbf{x}\right\|_2}.
\end{equation}
Similarly, solution to the  minimization problem given in  equation
(\ref{eq:dsmin})  is the proximal map of 
$\mathcal{N}_s(\lambda {\bf 1},  \cdot, p)$  applied on 
$\bar{\mathbf{d}}_s^{(k)} = 
\mathbf{D}_s^{\prime}{\bf E}^{(j)}\bar{\mathbf{s}}^{(k)} 
+ (1/\gamma){\bf w}_s^{(k)}$. Let 
$|||\cdot|||_t$  denote the operator that applies soft-thresholding
on the Eigen values of its matrix arguments and returns the resulting matrix.
Then the  proximal map of $\mathcal{N}_s(\lambda {\bf 1},  \cdot, p)$  
can be expressed as \cite{HS_13}
\begin{equation}
	\label{eq:admm_ds}
	\mathbf{d}_s^{(k+1)} = \sum_i^{N^2} \mathbf{Q}_i^T
	{\cal H_T}(\mathbf{Q}_i\bar{\mathbf{d}}_s^{(k)}, 
	\lambda /\gamma, p),
\end{equation}
where 
\begin{equation}
	{\cal H_T}(\mathbf{x}, t, p) = 
	\begin{cases}
		max(
		\|{\cal 
		S}(\mathbf{x})\|_F-t,0)\frac{\mathbf{x}}{\|\cal{S}(\mathbf{x})\|_F}, \;
		\mbox{for}\;\; p=2 \\
		{\cal S}^{-1}(|||{\cal S}(\mathbf{x})|||_t), \;\; \mbox{for}\;\; p =1
	\end{cases}
\end{equation}
Finally, the minimum defined in  equation (\ref{eq:d0min}) is the proximal map
of ${\cal B}({\bf d}_0)$   applied on 
$\bar{\mathbf{d}}_0^{(k)} = 
{\bf E}^{(j)}\bar{\mathbf{s}}^{(k)}
+ (1/\gamma){\bf w}_0^{(k)}$. This is denoted as
$\mathbf{d}_0^{(k+1)} = \mathcal{P}_u(\bar{\mathbf{d}}_0^{(k)} )$,
where $\mathcal{P}_u(\cdot)$ denotes the clipping of components of
the its vectors onto the range $[0,b]$, with  $b$  as the user-defined
upper bound.

The cost
$L_a(\bar{\mathbf{s}}, \mathbf{d}_f, \mathbf{d}_s, \mathbf{d}_0,
\mathbf{w}, \lambda )$ is differentiable w.r.t. $\bar{\bf s}$, and the 
minimization
defined in the equation can be obtained by equating  gradient to zero.
Let  $\mathbf{M} = 
\vecthe{{\bf D}_f^{\prime}}{{\bf D}_s^{\prime}}{\bf I}$
and  let
$\mathbf{d}^{(k+1)} = 
\vecthe{{\bf d}_f^{(k+1)}}{{\bf d}_s^{(k+1)}}{{\bf d}_0^{(k+1)}}$.  
Then, the  solution of the last minimization is given by the following equation,
\begin{equation}
	\label{eq:admmsksoln}
	\underbrace{\mathbf{E}^{(j)T}\mathbf{M}^T
		\mathbf{M}\mathbf{E}^{(j)}}_{{\mathbf A}_j}
	\mathbf{s}^{(k+1)} =  \mathbf{v}_{k+1},
\end{equation}
where  
$\mathbf{v}_{k+1} = 
\mathbf{E}^{(j)T}
\mathbf{M}^T\left(\mathbf{d}^{(k+1)}-(1/\gamma)\mathbf{w}^{(k+1)}\right)$.
Note that this equation has to be solved iteratively since $\mathbf{M}$ is 
composed
of non-circulant matrices.   We use conjugate 
gradient method for solving this problem. To speed-up, we use the inverse of
the following approximation of  the matrix  $\mathbf{A}_j$ in the equation
\eqref{eq:admmsksoln}, $\hat{\mathbf{A}}_j$, as the preconditioner:
\begin{equation}
	\label{eq:ahatj}
	\hat{\mathbf{A}}_j  = \mathbf{E}^{(j)T}\left(
	\mathbf{D}_f^T{\bf D}_f + \mathbf{D}_s^T{\bf D}_s + \mathbf{I}
	\right)\mathbf{E}^{(j)}
\end{equation}
All the matrices in the above product are circulant except $\mathbf{E}^{(j)}$.
However, the product, $\hat{\mathbf{A}}_j$, is circulant because of the special
structure of $\mathbf{E}^{(j)}$.  Hence, the preconditioning,  i.e.,  
multiplying 
by the inverse of $\hat{\mathbf{A}}_j$,  is equivalent to applying the inverse
of a discrete filter.  The following proposition gives the expression for this 
filter.
\begin{prop}
	\label{prop:filter}
	Let $u(z_1,z_2)$ be the z-transform of 
	$u(\r) = \frac{1}{64}[1\;4\;6\;4\;1]^T[1\;4\;6\;4\;1]$, and let   
	$u_j(z_1,z_2)=\prod_{i=0}^{j-1}u(z^{2^i}_1,z^{2^i}_2)$. Let
	$u_j(\r)$ be the inverse z-transform of $u_j(z_1,z_2)$. Then the filter 
	equivalent of
	$\hat{\mathbf{A}}_j$  is the $2^j$-fold decimation   of 
	$B(\r)= u_j(\r)*u_j(-\r)*[1+d_x(\r)*d_x(-\r)+d_y(\r)*d_y(-\r)
	+d_{xx}(\r)*d_{xx}(-\r)
	+d_{yy}(\r)*d_{yy}(-\r)+d_{xy}(\r)*d_{xy}(-\r)]$.
\end{prop}

Next, applying these ADMM steps described above for solving the minimization 
problem
of equation \eqref{eq:bcd2} is nearly identical except the fact that the 
up-sampling
matrix $\mathbf{E}$ is replaced by identity matrix because the cost is not 
defined
through up-sampling.  Here the size of the variable is the same as the size of 
the
measured image.   The initialization for ADMM iteration,  
$\bar{\mathbf{s}}^{(0)}$,
now simply comes from $s_{(k)}$,  which is the result of previous iteration of 
the 
BCD loop represented by equations \eqref{eq:bcd1} and \eqref{eq:bcd2}. 
Next,  the following proposition confirms the convergence of the block 
coordinate
descent algorithm specified by the equations \eqref{eq:bcd1} and 
\eqref{eq:bcd2}.
\begin{prop}
	\label{prop:bcd}
	The block coordinate descent method represented by the equations
	\eqref{eq:bcd1} and \eqref{eq:bcd2} with the problem of equation 
	\eqref{eq:bcd2} solved by ADMM method described above,  converges to a 
	local solution of the problem  \eqref{eq:corosa_cost_both}
	if  $J_{sa}(s_{(k+1)}, \beta_{(k+1)}, \tau, h, m)  <
	J_{sa}(s_{(k)}, \beta_{(k+1)}, \tau, h, m)$ and if
	there is no $s$ for which $F(s,h,m)$, 
	$ R_1(s) = \sum_{\bf r}  \|({\bf d}_1*s)({\bf r}) \|_2$, and 
	$R_2(s) = 
	\sum_{\bf r}  \|({\bf H}_2*s)({\bf r}) \|_2$ will have zero value 
	simultaneously.
\end{prop}

\section{Experimental results}
\label{sec:experiments}

For evaluating the restoration performance of the proposed COROSA approach, we 
considered deconvolution of Total Internal Reflection Fluorescence (TIRF) 
microscopy images and the reconstruction of Magnetic Resonance Images (MRI)  
from under-sampled Fourier data. These problems involve different measurement 
and noise models and are hence good candidates for evaluating the performance 
of the proposed approach alongside the state-of-the-art methods. 
We compare COROSA with second order TV (TV2) \cite{Scherzer_tv2_98}, 
Hessian-Schatten norm regularization (HS) \cite{HS_13}, combined order TV 
(COTV) \cite{Combined_order_TV} and  TGV2 \cite{TGV2} regularization methods. 
We also implemented the combined order TV formulation with Hessian-Schatten 
norm regularization replacing the original second order TV term, for the 
purpose of comparison. We refer to this method as Combined Order 
Hessian-Schatten (COHS) regularization. For the HS functional, we found that 
setting $p=1$  yielded the best performance. 
We also include the result of the multi-resolution loop represented by equations
\eqref{eq:mrbeta} and \eqref{eq:mrsi} in the comparison (without BCD iterations 
of
equations \eqref{eq:bcd1} and \eqref{eq:bcd2}).  We denote this by COROSA-I.
For objective comparison, we use Signal to Noise Ratio (SNR) 
\cite{bertero_inverse} 
and Structural Similarity Index (SSIM) \cite{ssim} scores. In tables,  we use 
COR. and
COR-I  to denote COROSA and COROSA-I respectively.

The smoothing parameter $\lambda$ was tuned for best performance in terms of 
SSIM and SNR  by using original reference images as done by most methods that 
focus on the design of regularization. In the case of COTV,  and COHS, 
additional 
tuning is required to fix the parameters determining the first and second order 
TV 
terms.  In this regard, we set the relative weights between first and second  
order 
derivatives so as to yield the lowest regularization functional cost. This 
ensures that 
only  $\lambda$ is required to be tuned using the reference images.

The spatial weight $\beta(\r)$ can be  determined
through optimization problem defined in 
\eqref{eq:mrbeta} and the corresponding result is given in Proposition
\ref{prop:beta}, with the parameter $\tau$ chosen to be a  scalar parameter.
However, we observed that it is advantageous to make $\tau$ spatially variant
for the following reasons: (i) in the regions of low intensity, it is 
advantageous 
to make $\beta(\r)$ less sensitive to variations in the relative magnitude of
first- and second-order derivatives, and hence $\tau$ has to be larger;
(ii) in the regions of high intensity, it is advantageous to make  $\beta(\r)$ 
more
sensitive to variations in the relative magnitude of
these terms, and hence $\tau$ has to be smaller.  
In short,  it is advantageous to make $\tau$ spatially variant and inversely
proportional to some approximate estimate of the required image, say 
$\bar{f}(\r)$.
We use the following strategy to get this approximate estimate:  
(i) for solving the problem given the equation  \eqref{eq:mrbeta}, we use
$f$ itself as the approximate estimate, i.e., we set $\bar{f}(\r)=f(\r)$.  
Next,  in the block-coordinate descent loop specified by the equations
\eqref{eq:bcd1}  and  \eqref{eq:bcd2},  we keep $\bar{f}(\r)$  fixed as the
image used to initialize the loop,  that is we set $\bar{f}(\r)=s_{(0)}(\r)$.
From $\bar{f}(\r)$, we compute $\tau(\r)$ as follows: 
we compute the image $\exp(-100*\bar{f}^{2}(\r))$ and then rescale it 
to the range $[0.01,100]$.  This scheme worked well  for all our test cases,
and hence we kept this scheme for determining $\tau(\r)$  in all test cases.

\begin{figure}[htbp]
	\centering 
	\includegraphics[width= \textwidth]{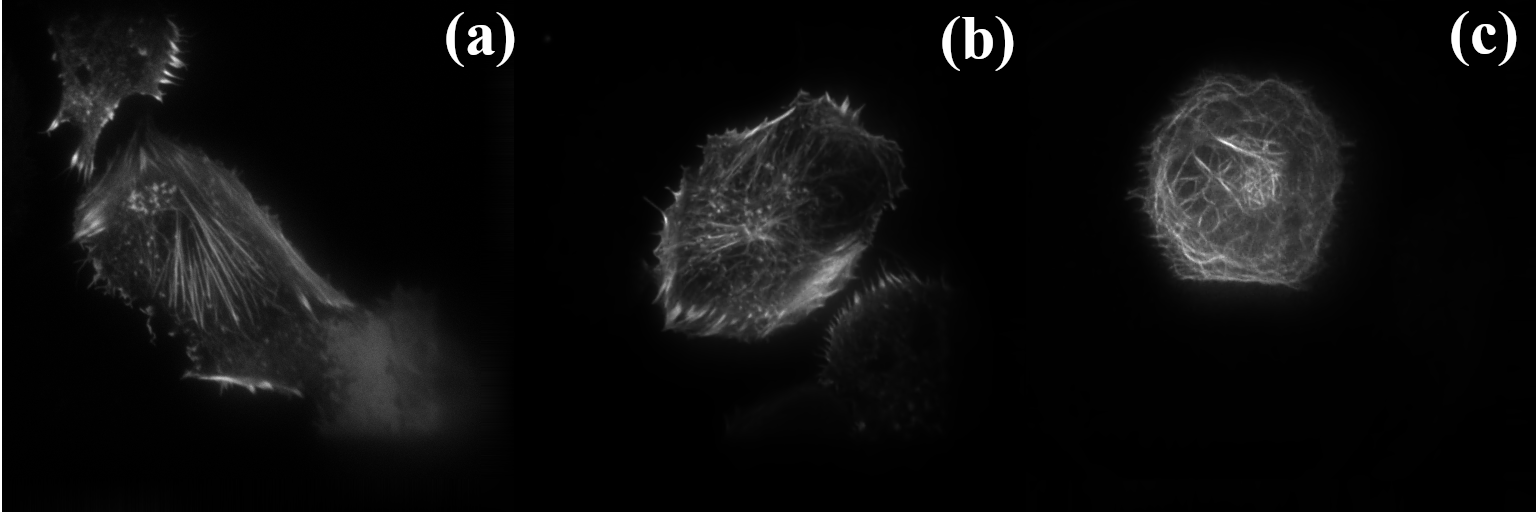}
	\caption{TIRF model images: (a) ActinSample1; (b) ActinSample2; (c) 
	TubulinSample.}
	\label{fig:tirf_source}
\end{figure}

\begin{figure}[htbp]
	\centering
	\includegraphics[width= 0.8\textwidth]{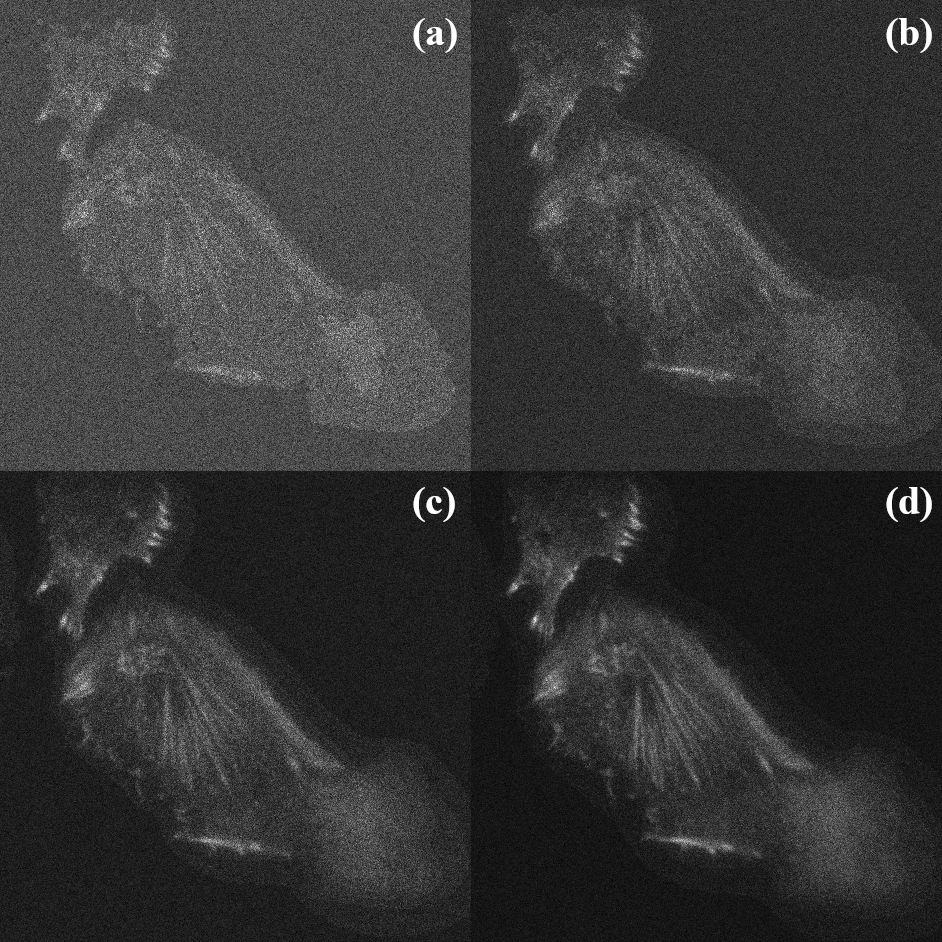}
	\caption{Blurred noisy images obtained from ActinSample1: (a) $\gamma_{p} = 
	5$; (b) $\gamma_{p} = 10$; (c) $\gamma_{p} = 20$; (d) $\gamma_{p} = 30$.}
	\label{fig:tirf_actin_images}
\end{figure}

In the first experiment,  we evaluate the proposed method for deconvolution of 
TIRF images. 
To obtain realistic ground-truth models,  we   measured TIRF images with 
negligible noise  and
deconvolved them by a simple inverse filtering.  The resulting images became 
the ground truth
models, which were used to simulate noisy blurred images for evaluating the 
proposed method
and for comparing it with  the existing methods. 
The  images for generating  ground-truth model   were acquired from samples 
containing
labelled  Actin and Tubulin. The Actin images   were acquired by staining with 
phalloidin-488 
and  an EMTB-mCherry transgene was used  to get the  Tubulin  image. Wavelength 
for excitation 
was 491nm for Actin samples and 561nm for Tubulin sample. 
All images were acquired using a
100x objective lens with numerical aperture 
1.45 NA.  The exposure time was set to 300ms,  and this was sufficient to 
get nearly noise-free
images with a quality that is adequate for using them as ground-truth models.
These are shown in Figure \ref{fig:tirf_source}.

\begin{figure}[htbp]
	\centering
	\includegraphics[width= 0.65\textwidth]{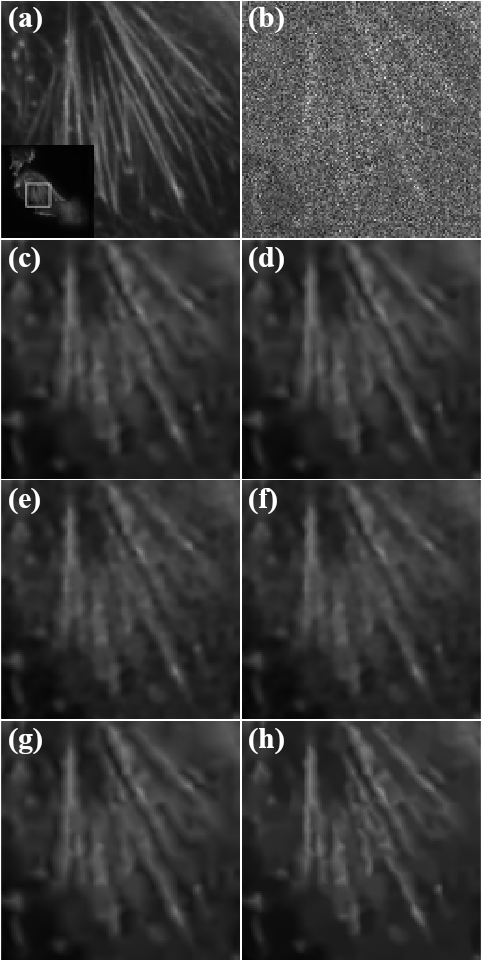}
	\caption{ActinSample1 Restoration: (a) Original Image; (b) Blurred Noisy 
	Image with $\gamma_{p} = 20$;
		(c) TV2; (d) HS; (e) COTV; (f) COHS; (g) COROSA-I; (h) COROSA.}
	\label{fig:tirf_actin_result1}
\end{figure}

For simulating  noisy blurred   images from ground-truth models,  we used 
blurring kernel  with a  bandwidth that is slightly lower than
the bandwidth of the system that measured the model images. 
This was done by setting NA = 1.4, while computing the required blurring kernel.
We used the following model to generate
test images: 
\begin{equation}
	m(\r) = {\mathcal P}[\gamma_{p}(h \ast s)(\r))] + \eta(\r)
\end{equation} 
Here $h(\r)$ represents the simulated 2D TIRF PSF with NA = 1.4,  and
with other parameters set to be identical to that of the system that measured 
the models. 
Next, ${\mathcal P}(\cdot)$ refers to Poisson process with $\gamma_{p}$ 
representing the scale factor for photon count, 
and $\eta$ is AWGN of variance $\sigma_{\eta}$.  We use this mixed noise model 
to make the test data
realistic, although we do not use the log-likelihood  functional of the mixed 
model in the formulation of our method.
For each reference  model, we set  $\sigma_{\eta} = 1$  and  generated four 
noisy blurred 
images with    
$\gamma_{p}$ = 5, 10, 20, and 30.
The four noisy images corresponding to ActinSample1 are shown in Figure 
\ref{fig:tirf_actin_images} as an example. 
It has to be pointed out 
that we did not include TGV in this experiment because of its poor performance, 
as was observed in 
\cite{sam-tv}.   In this regard, it is worthwhile to note that TGV has been 
used notably only for   MRI
reconstruction. The deconvolution results in terms of both SSIM and SNR scores 
are presented in 
Table \ref{table:1}. The scores show that the proposed COROSA outperforms other 
methods in most
cases, with the performance advantage significant when the noise is high. It is 
to be noted that 
TV2 gives slightly better scores than COROSA as the measurements become less 
noisy. 
This is due to the fact that  the effect of spatial adaptiveness is 
negligible,   and   the fact that,
TV2 can converge to the minimum more accurately because of its simplicity, 
which takes over the
advantage of spatial adaptiveness.    However, in such low-noise cases,  the 
difference in the score is
much lower than the advantage that COROSA has in noisy cases.  Figure 
\ref{fig:tirf_actin_result1} shows a set of restored images corresponding to 
ActinSample1 image and $\gamma_{p} = 20$. In terms of SSIM, the 
difference between TV2 and COROSA is 0.006 and difference in SNR is 0.2dB.  
However, as evident
from the displayed images,  there is a clear visual improvement in the result 
of COROSA.  
Another observation is that COROSA significantly outperforms COTV and COHS,  
because of the
spatial adaptivity. 
\begin{table}[htbp]
	\caption{Comparison of deconvolution results for TIRF images}
	\label{table:1}
	\centering
	\scalebox{1}{
		\begin{tabular}{|c|c|c|c|c|c|c|c|c|c|c|c|c|c|}
			\hline
			\multirow{2}{*}{IMG} & \multirow{2}{*}{\begin{tabular}[c]{@{}c@{}} 
			$\gamma_p$ \\\end{tabular}} & \multicolumn{6}{c|}{SSIM} & 
			\multicolumn{6}{c|}{SNR} \\ \cline{3-14}
			&  & TV2 & HS & COTV & COHS & COR-I & COR. & TV2 & HS & COTV & COHS 
			& COR-I & COR. \\ \hline
			\multirow{4}{*}{Act.1} & 5 & .725 & .731 & .502 & .502 & .728 & 
			\textbf{.762} & 10.53 & 10.58 & 9.37 & 9.37 & 10.58 & 
			\textbf{10.97} \\ \cline{2-14} 
			& 10 & .794 & .793 & .679 & .679 & .792 & \textbf{.807} & 13.44 & 
			13.45 & 12.88 & 12.89 & 13.40 & \textbf{13.82} \\ \cline{2-14} 
			& 20 & .853 & .849 & .806 & .807 & .848 & \textbf{.853} & 15.62 & 
			15.64 & 15.38 & 15.43 & 15.59 & \textbf{15.90} \\ \cline{2-14} 
			& 30 & \textbf{.890} & .888 & .867 & .867 & .887 & .890 & 16.88 & 
			16.91 & 16.81 & 16.81 & 16.81 & \textbf{17.08} \\ \hline
			\multirow{4}{*}{Act.2} & 5 & .735 & .744 & .516 & .516 & .741 & 
			\textbf{.780} & 12.03 & 12.06 & 10.74 & 10.74 & 12.05 & 
			\textbf{12.34} \\ \cline{2-14} 
			& 10 & .828 & .829 & .712 & .712 & .829 & \textbf{.844} & 14.59 & 
			14.58 & 13.96 & 13.97 & 14.58 & \textbf{14.74} \\ \cline{2-14} 
			& 20 & \textbf{.883} & .881 & .839 & .839 & .880 & .882 & 
			\textbf{16.58} & 16.56 & 16.38 & 16.39 & 16.54 & 16.55 \\ 
			\cline{2-14} 
			& 30 & \textbf{.904} & .900 & .880 & .880 & .897 & .899 & 
			\textbf{17.69} & 17.66 & 17.58 & 17.59 & 17.60 & 17.57 \\ \hline
			\multirow{4}{*}{Tub.1} & 5 & .769 & .784 & .491 & .491 & .780 & 
			\textbf{.827} & 13.42 & 13.60 & 11.30 & 11.30 & 13.57 & 
			\textbf{13.92} \\ \cline{2-14} 
			& 10 & .830 & .832 & .685 & .685 & .830 & \textbf{.844} & 15.70 & 
			15.71 & 14.76 & 14.76 & 15.69 & \textbf{15.86} \\ \cline{2-14} 
			& 20 & \textbf{.868} & .864 & .806 & .806 & .863 & .862 & 17.46 & 
			17.52 & 17.17 & 17.18 & 17.50 & \textbf{17.55} \\ \cline{2-14} 
			& 30 & \textbf{.881} & .877 & .850 & .850 & .875 & .875 & 18.27 & 
			\textbf{18.29} & 18.07 & 18.11 & 18.23 & 18.25 \\ \hline
		\end{tabular}
	}
\end{table}

%

\begin{figure}[htbp]
	\begin{minipage}{0.53 \textwidth}
		\centering
			\includegraphics[width= \textwidth]{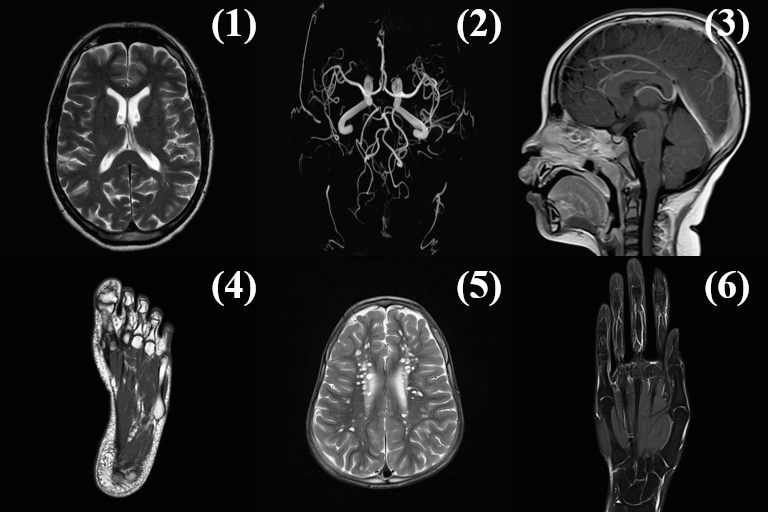}
			\caption{MRI Reference Images}
			\label{fig:mri_source}
	\end{minipage}
	\begin{minipage}{0.46 \textwidth}
			\centering
			\includegraphics[width= 0.8\textwidth]{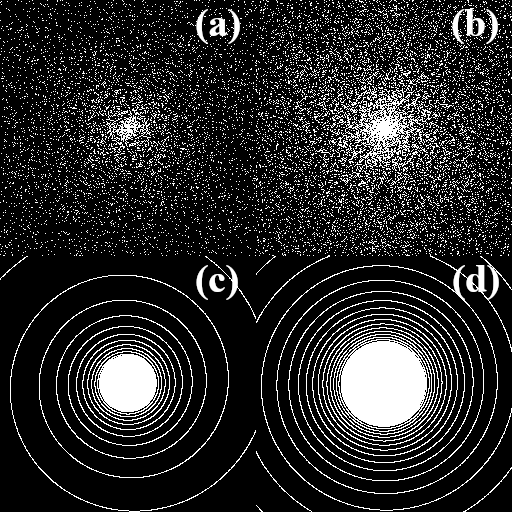}
			\caption{MRI Trajectories. Random: (a) 10\% (b) 20\%, Spiral: (c) 
		10\% (d) 
			20\%}
			\label{fig:mri_sampling}
	\end{minipage}
\end{figure}

In the second experiment, we considered reconstruction of MRI images from 
undersampled k-space measurements. The source MRI images are shown in Figure 
\ref{fig:mri_source}.  It has to be mentioned that some of  these reference 
images 
have been obtained  using traditional reconstruction methods and hence contain 
artifacts. However,  this does not
affect the validity of our comparisons done here, because,  all the methods 
evaluated in this paper including the
proposed method remove these artifacts.  Hence, we focus on how well the actual 
structures are reproduced by
various methods in the presence of noise and undersampling, and ignore  the 
removal  of artifacts present in the reference images. For sampling, we 
generated 
random and spiral trajectories using the MATLAB code given by Chauffert et al. 
\cite{MRI_Chauffert}. In addition, we 
modified the spiral trajectory by filling in low frequency region, since the 
default spiral had  low sample density in the low frequency region. Both 
sampling trajectories with 10\% and 20 \% densities are shown in Figure 
\ref{fig:mri_sampling}. 
For simulating the thermal noise,  we added white Gaussian noise to the k-space 
data.  To control the noise,
we use the strategy of \cite{MRI_Bresler}: 
the Gaussian variance is adjusted such that, if added to both real and 
imaginary parts of full Fourier transforms
of the image,   it results in  PSNR of 10dB and 20dB  upon the Fourier 
inversion.  
We generated six sample sets from  each MRI image as follows:  for each 
sampling trajectory, we generated
20\% sample set with two noise levels (10dB and 20dB),  and  10\% sample set 
with one noise level  (20dB). 
This make a total of 36 sample sets.   The comparisons in terms of   SSIM and 
SNR scores for all cases are given in Table \ref{table:2}.

\begin{figure*}[htbp]
	\centering
	\includegraphics[width= \textwidth]{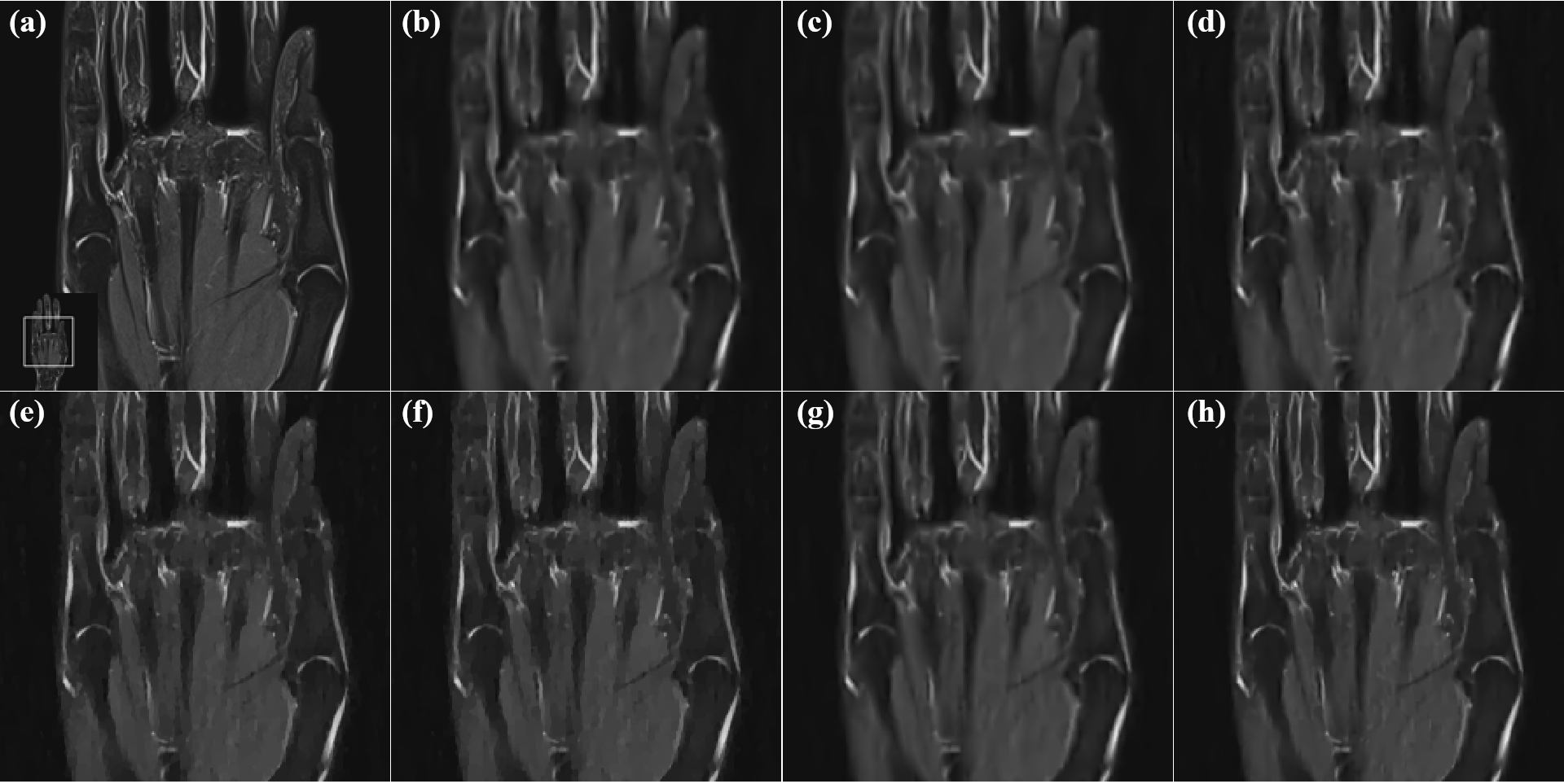}
	\caption{MRI 6 Restoration (10\% Random Samples, 20dB measurement SNR) (a) 
	Reference (b) TV2 (c) HS (d) TGV (e) COTV (f) COHS (g) COROSA-I (h) COROSA}
	\label{fig:mri_6_sr_1_sig_20_ssim}
\end{figure*}

\begin{figure*}[htbp]
	\centering
	\includegraphics[width= 
	0.98\textwidth]{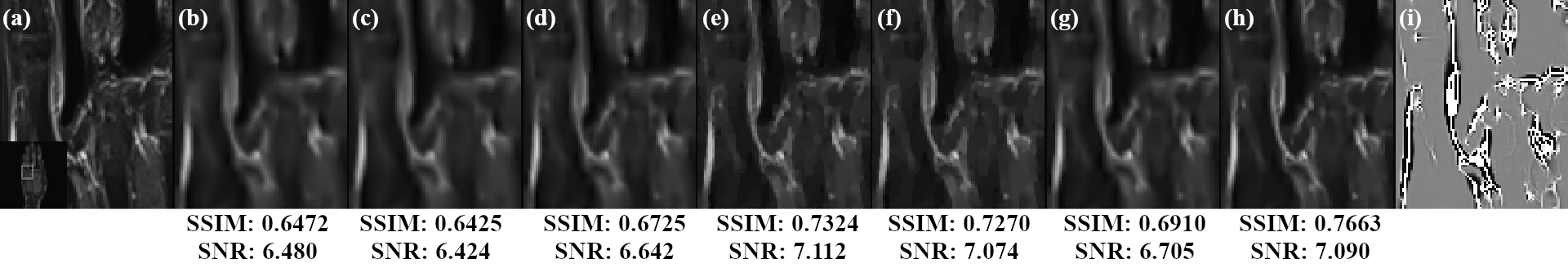}
	\caption{Comparison of Selected Region in MRI 6 Restoration (from Figure 
	\ref{fig:mri_6_sr_1_sig_20_ssim}) (a) Reference (b) TV2 (c) HS (d) TGV (e) 
	COTV (f) COHS (g) COROSA-I (h) COROSA (i) corresponding adaptive weight, 
	$\beta(\r)$}
	\label{fig:mri_6_sr_1_sig_20_ssim2}
\end{figure*}

\begin{figure*}[htbp]
	\centering
	\includegraphics[width= 
	0.98\textwidth]{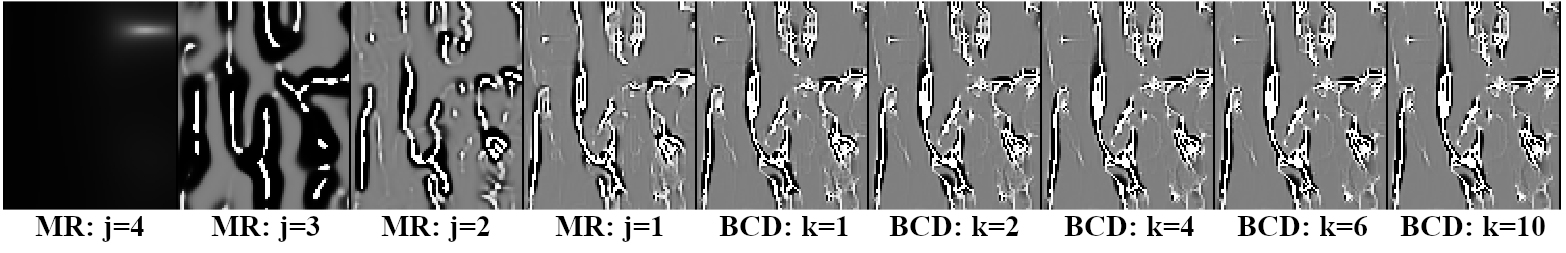}
	\caption{ Evolution of adaptive weight $\beta({\bf r})$ corresponding to 
	Figure \ref{fig:mri_6_sr_1_sig_20_ssim2}: ``MR, j"  denotes the adaptive
		weight at $j$th level of multiresolution loop. ``BCD, k"  denotes the 
		adaptive weight at $k$th iteration of block coordinate descent loop.}	
		\label{fig:mri_6_sr_1_sig_20_weight}
\end{figure*}

The reconstruction scores in Table \ref{table:2} show that, in most cases, 
the proposed COROSA approach is better than all competitive methods, 
with COTV and COHS being the nearest in terms of scores. 
In terms of SSIM,  COROSA is always better, and in terms of SNR,
COROSA is better in most of the cases.
In general, combined order methods perform better in 
MRI reconstruction, when compared to TIRF restoration. 
Figure \ref{fig:mri_6_sr_1_sig_20_ssim} shows the reconstruction results 
for MRI 6 image with 10\% sampling and 20dB PSNR samples. It is clear 
from the figure that COROSA suppresses background artifacts with the 
help of spatial adaptation better than TV2 and HS.  At the same time,
COROSA also  avoids 
staircase artifacts that is normally caused if TV1  is used without spatial
adaptation as present  in COTV and COHS reconstructions. A selected region of 
this 
reconstruction is shown in Figure \ref{fig:mri_6_sr_1_sig_20_ssim2}
to further emphasize this fact, along with the relative weight  $\beta(\r)$.
In the figure, white regions correspond to locations where   first order TV 
term is dominant,
and black regions correspond to locations where   second order TV term is 
dominant.
Grey regions represent locations where both orders are weighted equally.  
It is clear from the figure that spatial weights follow the local intensity 
structure, 
leading to     the elimination  of staircase artifacts. 
On the other hand, such artifacts are evident in COTV 
reconstruction, which uses similar cost function and optimization framework, 
albeit 
with constant global relative weights.  Recall that SSIM of COROSA for this
selected region is 0.7663 and that of COTV is 0.7324.  
Figure \ref{fig:mri_6_sr_1_sig_20_weight} shows the evolution of adaptive 
weight $\beta({\bf r})$  across different stages of multi-resolution  and
block coordinate descent iterations. It can be noted that $\beta({\bf r})$ 
follows
fine variations in the relative magnitudes of first- and second-order 
derivatives
that are not obvious from the visual inspection of   the displayed intensity 
images.

\begin{figure*}[htbp]
	\centering
	\includegraphics[width= 
	\textwidth]{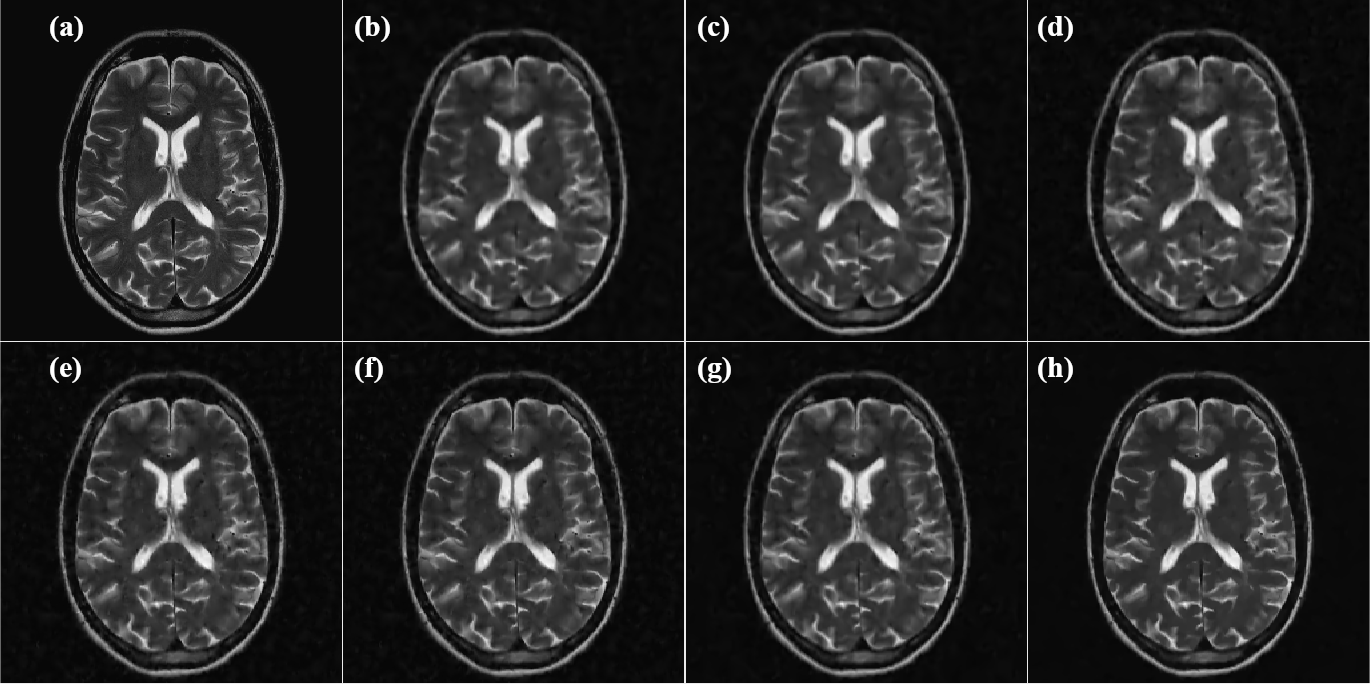}
	\caption{MRI 1 Restoration (10\% Random Samples, 20dB) (a) Reference (b) 
	TV2 (c) HS (d) TGV (e) COTV (f) COHS (g) COROSA-I (h) COROSA}
	\label{fig:mri_1_sr_1_sig_20_ssim1}
\end{figure*}

\begin{figure*}[htbp]
	\centering
	\includegraphics[width= 
	0.98\textwidth]{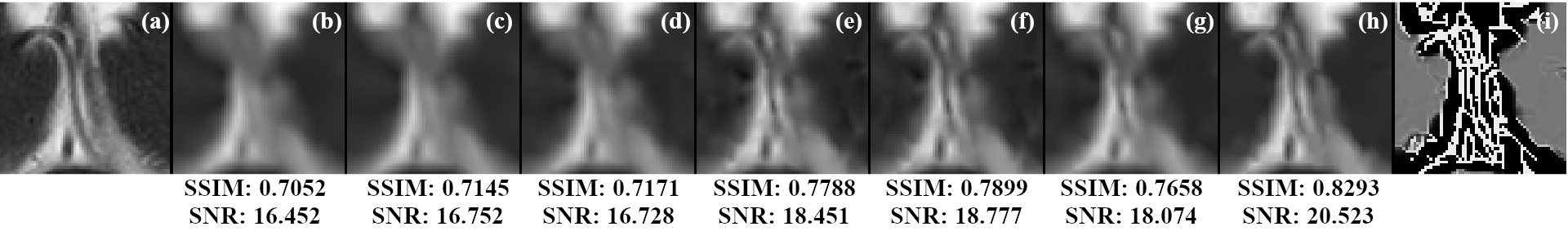}
	\caption{Comparison of Selected Region in MRI 1 Restoration (from Figure 
	\ref{fig:mri_1_sr_1_sig_20_ssim1}) 
		(a) Reference (b) TV2 (c) HS (d) TGV (e) COTV (f) COHS (g) COROSA-I (h) 
		COROSA (i) 
		Corresponding adaptive weight $\beta(\r)$}
	\label{fig:mri_1_sr_1_sig_20_ssim2}
\end{figure*}

We have also given the reconstruction results 
for MRI 1 image with 10\% sampling and 20dB PSNR samples in Figure
\ref{fig:mri_1_sr_1_sig_20_ssim1},  confirming  that the visual improvement 
with 
COROSA is consistent with the higher scores. Further, COTV and COHS 
introduce prominent artifacts in the  reconstruction as seen in zoomed-in
view of the same set of images displayed in Figure
\ref{fig:mri_1_sr_1_sig_20_ssim2} along with $\beta(\r)$. It is also clear that 
these artifacts
are not the remnants of any artifacts from the original model,  but were 
produced
by COTV and COHS methods.  On the other hand,  COROSA results do not have
any artifacts and the adaptive weights clearly pick out image boundaries from 
uniform intensity regions. 
There are two exceptions in the results that do not confer to the pattern 
exhibited    by
other test cases. In the first case, COTV was giving  
better SNR than COROSA, with comparable SSIM scores for reconstruction of MRI 
1, MRI 2 
and MRI 4 images from spiral trajectory. This is because of the modification in 
the spiral trajectory, where
low frequency regions are filled and hence the advantage of multiresolution 
framework utilized in 
COROSA   becomes less significant.
In the second case,  COTV was giving higher SNR score for MRI 4 image 
reconstruction from 
random sampling, while having lower SSIM score when compared to COROSA. 
When we evaluated the corresponding
results visually, we found that the reference MRI 4 image itself has block-like
piecewise constant regions. Because of this, COTV results have higher SNR
compared to COROSA  because of the fact that COROSA keeps a minimum amount
of second-order smoothing even in the piece-wise constant regions.  However, in 
terms
of SSIM score and visual quality, the difference is insignificant.
Next, it is worthwhile to note that, TGV has the ability to be spatially 
adaptive  because of the 
auxiliary variable, $\mathbf{p}$, and it also retains convexity, which makes it 
quite attractive. However,
our method outperforms TGV significantly. Further,  surprisingly,  even basic 
non-adaptive
methods such as COTV and COHS outperform TGV in many test cases.  A possible 
reason that
we inferred based on some  reconstructions trials,  is that,  the inferior 
performance of TGV is due
to  the lack of efficient optimization method to handle the auxiliary variable. 
Specifically, the 
convergence of all known optimization methods proposed for TGV  is highly 
dependent on the 
value of the smoothing parameters $\alpha_1$, $\alpha_2$ and $\lambda$. This 
leads to the 
inferior  performance of TGV although it  is based on rich and elegant 
mathematical formulation. 

\begin{table*}[htbp]
	\caption{Comparison of MRI reconstruction results}
	\label{table:2}
	\centering
	\scalebox{0.9}{
		\begin{tabular}{|c|c|c|c|c|c|c|c|c|c|c|c|c|c|c|c|c|c|}
			\hline
			\multirow{2}{*}{IMG} & \multirow{2}{*}{Traj.} & \multirow{2}{*}{SR} 
			& \multirow{2}{*}{\begin{tabular}[c]{@{}c@{}}I/P\\ 
			PSNR\end{tabular}} & \multicolumn{7}{c|}{SSIM} & 
			\multicolumn{7}{c|}{SNR} \\ \cline{5-18} 
			&  &  &  & TV2 & HS & TGV & COTV & COHS & COR-I & COR. & TV2 & HS & 
			TGV & COTV & COHS & COR-I & COR. \\ \hline
			\multirow{6}{*}{MRI\_1} & \multirow{3}{*}{Spiral} & 
			\multirow{2}{*}{.2} & 10 & .972 & .976 & .979 & \textbf{.983} & 
			.982 & .976 & \textbf{.983} & 24.71 & 25.78 & 26.68 & 
			\textbf{28.26} & 28.16 & 25.72 & 28.07 \\ \cline{4-18} 
			&  &  & 20 & .972 & .976 & .979 & \textbf{.983} & .982 & .976 & 
			\textbf{.983} & 24.71 & 25.78 & 26.68 & \textbf{28.27} & 28.19 & 
			25.73 & 28.12 \\ \cline{3-18} 
			&  & .1 & 20 & .955 & .955 & .959 & .964 & .964 & .956 & 
			\textbf{.966} & 22.07 & 22.14 & 22.86 & \textbf{24.20} & 24.12 & 
			22.35 & 24.09 \\ \cline{2-18} 
			& \multirow{3}{*}{Random} & \multirow{2}{*}{.2} & 10 & .902 & .927 
			& .927 & .926 & .934 & .946 & \textbf{.962} & 19.39 & 20.74 & 19.72 
			& 22.80 & 22.98 & 21.84 & \textbf{24.35} \\ \cline{4-18} 
			&  &  & 20 & .902 & .927 & .927 & .926 & .934 & .946 & 
			\textbf{.962} & 19.39 & 20.73 & 19.75 & 22.81 & 22.99 & 21.85 & 
			\textbf{24.35} \\ \cline{3-18} 
			&  & .1 & 20 & .807 & .811 & .793 & .810 & .810 & .839 & 
			\textbf{.905} & 14.42 & 14.73 & 14.38 & 16.38 & 16.38 & 15.71 & 
			\textbf{18.52} \\ \hline
			\multirow{6}{*}{MRI\_2} & \multirow{3}{*}{Spiral} & 
			\multirow{2}{*}{.2} & 10 & .990 & .993 & .995 & \textbf{.998} & 
			\textbf{.998} & .994 & \textbf{.998} & 22.68 & 24.53 & 26.06 & 
			\textbf{30.66} & \textbf{30.66} & 24.70 & 30.28 \\ \cline{4-18} 
			&  &  & 20 & .990 & .993 & .995 & \textbf{.998} & \textbf{.998} & 
			.994 & \textbf{.998} & 22.68 & 24.54 & 26.06 & \textbf{30.77} & 
			\textbf{30.77} & 24.71 & 30.45 \\ \cline{3-18} 
			&  & .1 & 20 & .978 & .978 & .982 & \textbf{.993} & .993 & .982 & 
			.992 & 18.95 & 19.09 & 20.20 & \textbf{24.40} & \textbf{24.40} & 
			19.93 & 23.58 \\ \cline{2-18} 
			& \multirow{3}{*}{Random} & \multirow{2}{*}{.2} & 10 & .936 & .963 
			& .965 & .992 & .992 & .983 & \textbf{.995} & 16.69 & 18.87 & 17.69 
			& 27.49 & 26.60 & 20.76 & \textbf{27.54} \\ \cline{4-18} 
			&  &  & 20 & .936 & .963 & .957 & .992 & .992 & .983 & 
			\textbf{.995} & 16.69 & 18.88 & 16.65 & 26.65 & 26.65 & 20.77 & 
			\textbf{27.60} \\ \cline{3-18} 
			&  & .1 & 20 & .853 & .856 & .846 & .887 & .904 & .895 & 
			\textbf{.959} & 11.96 & 12.14 & 11.77 & 15.26 & 15.48 & 13.44 & 
			\textbf{17.38} \\ \hline
			\multirow{6}{*}{MRI\_3} & \multirow{3}{*}{Spiral} & 
			\multirow{2}{*}{.2} & 10 & .940 & .948 & .952 & .961 & .961 & .948 
			& \textbf{.962} & 21.50 & 22.61 & 23.12 & 24.55 & 24.55 & 22.67 & 
			\textbf{24.56} \\ \cline{4-18} 
			&  &  & 20 & .940 & .948 & .952 & .961 & .961 & .949 & 
			\textbf{.962} & 21.50 & 22.62 & 23.12 & 24.57 & 24.57 & 22.67 & 
			\textbf{24.58} \\ \cline{3-18} 
			&  & .1 & 20 & .879 & .882 & .879 & .891 & .895 & .880 & 
			\textbf{.898} & 18.00 & 18.18 & 18.33 & \textbf{19.58} & 
			\textbf{19.58} & 18.33 & 19.44 \\ \cline{2-18} 
			& \multirow{3}{*}{Random} & \multirow{2}{*}{.2} & 10 & .750 & .784 
			& .760 & .790 & .790 & .804 & \textbf{.864} & 14.88 & 16.04 & 15.45 
			& 17.32 & 17.22 & 16.72 & \textbf{18.33} \\ \cline{4-18} 
			&  &  & 20 & .743 & .780 & .753 & .783 & .783 & .796 & 
			\textbf{.854} & 14.81 & 15.92 & 15.31 & 16.96 & 16.93 & 16.49 & 
			\textbf{17.98} \\ \cline{3-18} 
			&  & .1 & 20 & .498 & .503 & .486 & .547 & .562 & .547 & 
			\textbf{.651} & 8.64 & 8.79 & 8.73 & 9.71 & 11.02 & 9.43 & 
			\textbf{11.51} \\ \hline
			\multirow{6}{*}{MRI\_4} & \multirow{3}{*}{Spiral} & 
			\multirow{2}{*}{.2} & 10 & .992 & .994 & .990 & \textbf{.996} & 
			\textbf{.996} & .994 & \textbf{.996} & 28.60 & 29.92 & 28.13 & 
			\textbf{33.22} & 32.66 & 30.12 & 32.65 \\ \cline{4-18} 
			&  &  & 20 & .992 & .994 & .990 & \textbf{.996} & \textbf{.996} & 
			.994 & \textbf{.996} & 28.60 & 29.94 & 28.13 & \textbf{33.28} & 
			32.72 & 30.14 & 32.73 \\ \cline{3-18} 
			&  & .1 & 20 & .983 & .982 & .984 & \textbf{.989} & .988 & .984 & 
			\textbf{.989} & 24.69 & 24.60 & 25.60 & \textbf{28.01} & 27.56 & 
			25.42 & 27.21 \\ \cline{2-18} 
			& \multirow{3}{*}{Random} & \multirow{2}{*}{.2} & 10 & .971 & .977 
			& .978 & .978 & .979 & .980 & \textbf{.985} & 24.58 & 26.19 & 26.22 
			& \textbf{30.66} & 30.15 & 27.38 & 30.06 \\ \cline{4-18} 
			&  &  & 20 & .970 & .976 & .978 & .978 & .980 & .980 & 
			\textbf{.985} & 24.49 & 26.10 & 26.13 & \textbf{30.84} & 30.20 & 
			27.30 & 30.13 \\ \cline{3-18} 
			&  & .1 & 20 & .928 & .927 & .920 & .936 & .936 & .942 & 
			\textbf{.961} & 19.48 & 19.48 & 19.37 & 21.72 & 21.72 & 20.54 & 
			\textbf{22.54} \\ \hline
			\multirow{6}{*}{MRI\_5} & \multirow{3}{*}{Spiral} & 
			\multirow{2}{*}{.2} & 10 & .958 & .964 & .968 & .973 & .973 & .963 
			& \textbf{.974} & 21.00 & 21.98 & 22.75 & 24.28 & 24.28 & 21.95 & 
			\textbf{24.35} \\ \cline{4-18} 
			&  &  & 20 & .958 & .964 & .969 & .974 & .974 & .963 & 
			\textbf{.975} & 21.00 & 21.98 & 22.82 & 24.31 & 24.31 & 21.96 & 
			\textbf{24.38} \\ \cline{3-18} 
			&  & .1 & 20 & .922 & .923 & .926 & .936 & .936 & .927 & 
			\textbf{.938} & 18.16 & 18.19 & 18.50 & \textbf{19.63} & 
			\textbf{19.63} & 18.59 & 19.48 \\ \cline{2-18} 
			& \multirow{3}{*}{Random} & \multirow{2}{*}{.2} & 10 & .856 & .886 
			& .888 & .896 & .902 & .914 & \textbf{.951} & 17.08 & 18.11 & 17.47 
			& 19.89 & 19.89 & 18.86 & \textbf{21.60} \\ \cline{4-18} 
			&  &  & 20 & .851 & .880 & .880 & .890 & .894 & .912 & 
			\textbf{.954} & 16.97 & 18.02 & 17.43 & 19.88 & 19.88 & 18.89 & 
			\textbf{21.91} \\ \cline{3-18} 
			&  & .1 & 20 & .737 & .736 & .732 & .763 & .759 & .779 & 
			\textbf{.871} & 13.84 & 13.88 & 13.43 & 15.09 & 14.83 & 14.59 & 
			\textbf{16.16} \\ \hline
			\multirow{6}{*}{MRI\_6} & \multirow{3}{*}{Spiral} & 
			\multirow{2}{*}{.2} & 10 & .983 & .987 & .984 & .994 & .994 & .987 
			& \textbf{.995} & 19.36 & 20.98 & 19.96 & 25.89 & 25.89 & 21.14 & 
			\textbf{27.25} \\ \cline{4-18} 
			&  &  & 20 & .984 & .987 & .984 & .994 & .994 & .987 & 
			\textbf{.996} & 19.36 & 20.98 & 19.96 & 25.96 & 25.96 & 21.14 & 
			\textbf{27.43} \\ \cline{3-18} 
			&  & .1 & 20 & .972 & .972 & .977 & .984 & .984 & .975 & 
			\textbf{.986} & 17.15 & 17.18 & 18.62 & 20.96 & 20.96 & 17.76 & 
			\textbf{21.21} \\ \cline{2-18} 
			& \multirow{3}{*}{Random} & \multirow{2}{*}{.2} & 10 & .956 & .969 
			& .967 & .978 & .978 & .976 & \textbf{.989} & 16.61 & 18.26 & 17.00 
			& 22.09 & 22.09 & 18.95 & \textbf{23.84} \\ \cline{4-18} 
			&  &  & 20 & .952 & .965 & .968 & .976 & .976 & .976 & 
			\textbf{.990} & 16.78 & 18.54 & 17.22 & 22.42 & 22.42 & 19.18 & 
			\textbf{24.35} \\ \cline{3-18} 
			&  & .1 & 20 & .896 & .896 & .892 & .911 & .915 & .925 & 
			\textbf{.954} & 12.82 & 12.89 & 13.09 & 15.30 & 15.30 & 13.83 & 
			\textbf{16.67} \\ \hline
		\end{tabular}
	}
\end{table*}

\begin{figure}[htbp]
	\centering
	\includegraphics[width= 0.8\textwidth]{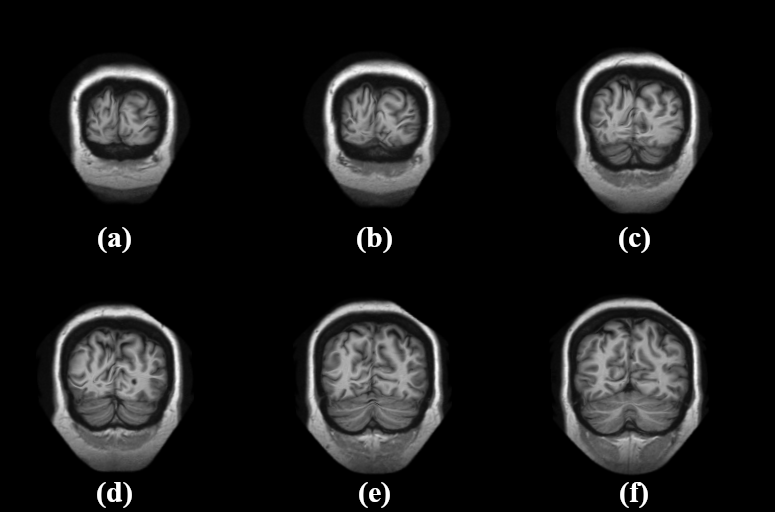}
	\caption{MRI Reference Images from DAGAN test dataset:
		(a) I1 (b) I2 (c) I3 (d) I4 (e) I5 (f) I6}
	\label{fig:dagan_source}
\end{figure}

Recently, there has been significant interest in deep learning based image 
restoration \cite{DAGAN, Deep_MRI_1, Deep_MRI_2, Deep_MRI_3} and dictionary 
learning methods \cite{MRI_Bresler, DL_Caballero14, DL_mubarakhigher}.  
Although  these methods require training samples, unlike our method which
does not need any training, we make comparison with representative methods from 
these categories to obtain
a perspective.  We  chose the method of  Deep de-aliasing GAN (DAGAN) proposed 
by Yang et al. \cite{DAGAN}  and the 
dictionary learning method of  Ravishankar et al. 
\cite{MRI_Bresler} for this purpose,  and  utilized the code provided by the  
authors.  For DAGAN we chose
the best Pixel-Frequency-Perceptual-GAN-Refinement (PFPGR) topology of the 
network for comparison, and also 
used the same dataset used by the authors with identical settings.  Since our 
method and the dictionary
learning methods  do not require  training samples,  we evaluated DAGAN with 
varying number of  training images: we evaluated this method with 10$\%$, 
50$\%$ and 100$\%$  of the 15972 training images
used by the authors. We selected 6 images from the set of 50 test images 
provided by the authors for
evaluation. These images are shown  in Figure \ref{fig:dagan_source}. 

\begin{figure}[htbp]
	\centering
	\includegraphics[width= 
	0.8\textwidth]{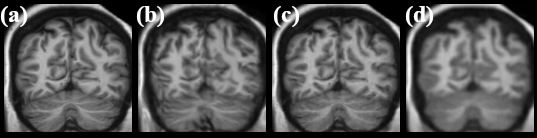}
	\caption{Selected Portion of MRI Reconstruction  (20 \% Noiseless samples): 
	(a) Original Image I6 (b) DAGAN (50\% of training set) result (c) COROSA 
	result (d) DL-MRI result}
	\label{fig:dagan_noiseless_20per}
\end{figure}

As the first part of this experiment, we tested reconstruction from 
under-sampled
data without noise 
using random trajectory. It has to be noted that DAGAN was not designed to take
complex measurements. 
Instead, it takes the absolute of the inverse Fourier transform of the 
zero-filled k-space values.
To eliminate this disadvantage for DAGAN,   we made the random sampling 
symmetric
in Fourier domain, yielding real images. The reconstruction results for 10\% 
and 20\%
sampling ratios are given in Table \ref{table:3}. As the results indicate, 
COROSA performs
better than DAGAN and DL-MRI consistently with $20\%$ samples,  in terms of both
SNR and SSIM. With $10\%$ sampling, COROSA performs better than the other 
methods
in most cases; however, in some cases, DAGAN has higher SNR scores than COROSA
although     COROSA is still better in terms of SSIM scores. This is
due to the fact  that DAGAN reconstructions have artifacts, and SSIM score is 
more sensitive
to artifacts than SNR measure.
Further, on visual inspection, we found that DL-MRI reconstructions are
over-smoothed, especially the reconstruction with 10\% samples. We also noted 
that
the reconstructions 
using COROSA  do not create artifacts in all test cases including the cases 
where DAGAN has better SNR, as seen in Figure
\ref{fig:dagan_noiseless_20per}. On the other hand,  DAGAN reconstruction shows 
spurious  structures introduced during reconstruction.  
The main point that we want to emphasize in this regard is that COROSA is able 
to adapt to image structure, due to the robust initialization from 
multi-resolution framework.
In summary, COROSA outperforms both DAGAN and DL-MRI in all cases in terms of 
SSIM score.

\begin{table}[ht]
	\caption{Comparison of MRI reconstruction results with DAGAN and DL-MRI 
		(Noiseless Measurements)}
	\label{table:3}
	\centering
	\scalebox{1}{
		\begin{tabular}{|c|c|c|c|c|c|c|c|c|c|c|c|}
			\hline
			\multirow{3}{*}{IMG} & \multirow{3}{*}{SR} & 
			\multicolumn{5}{c|}{SNR} & \multicolumn{5}{c|}{SSIM} \\ 
			\cline{3-12} 
			&  & \multirow{2}{*}{\begin{tabular}[c]{@{}c@{}}DL-\\ 
					MRI\end{tabular}} & \multicolumn{3}{c|}{DAGAN} & 
			\multirow{2}{*}{COR.} & 
			\multirow{2}{*}{\begin{tabular}[c]{@{}c@{}}DL-\\ MRI\end{tabular}} 
			& \multicolumn{3}{c|}{DAGAN} & \multirow{2}{*}{COR.} \\ \cline{4-6} 
			\cline{9-11}
			&  &  & 10\% & 50\% & 100\% &  &  & 10\% & 50\% & 100\% &  \\ \hline
			\multirow{2}{*}{I1} & .1 & 18.29 & 27.71 & \textbf{28.85} & 28.52 & 
			28.30 & .722 & .953 & .964 & .961 & \textbf{.991} \\ \cline{2-12} 
			& .2 & 22.73 & 31.23 & 32.35 & 31.70 & \textbf{37.78} & .949 & .971 
			& .978 & .975 & \textbf{.999} \\ \hline
			\multirow{2}{*}{I2} & .1 & 18.13 & 27.25 & 28.18 & \textbf{28.22} & 
			27.44 & .735 & .949 & .958 & .960 & \textbf{.989} \\ \cline{2-12} 
			& .2 & 22.20 & 30.66 & 32.04 & 31.28 & \textbf{36.68} & .944 & .967 
			& .976 & .972 & \textbf{.998} \\ \hline
			\multirow{2}{*}{I3} & .1 & 18.40 & 26.88 & 27.70 & \textbf{27.81} & 
			24.98 & .775 & .940 & .950 & .950 & \textbf{.978} \\ \cline{2-12} 
			& .2 & 22.09 & 29.87 & 30.67 & 30.59 & \textbf{34.62} & .934 & .963 
			& .969 & .969 & \textbf{.997} \\ \hline
			\multirow{2}{*}{I4} & .1 & 17.21 & 25.96 & 26.95 & \textbf{27.15} & 
			23.72 & .770 & .933 & .944 & .946 & \textbf{.973} \\ \cline{2-12} 
			& .2 & 21.93 & 28.56 & 29.29 & 29.27 & \textbf{32.92} & .935 & .956 
			& .964 & .962 & \textbf{.996} \\ \hline
			\multirow{2}{*}{I5} & .1 & 16.89 & 24.67 & \textbf{25.88} & 25.67 & 
			22.77 & .750 & .918 & .931 & .932 & \textbf{.963} \\ \cline{2-12} 
			& .2 & 21.90 & 28.56 & 29.45 & 29.14 & \textbf{31.98} & .928 & .954 
			& .961 & .960 & \textbf{.994} \\ \hline
			\multirow{2}{*}{I6} & .1 & 16.92 & 24.32 & \textbf{25.50} & 25.40 & 
			22.37 & .751 & .912 & .925 & .927 & \textbf{.958} \\ \cline{2-12} 
			& .2 & 21.58 & 27.91 & 28.65 & 28.46 & \textbf{31.09} & .924 & .948 
			& .956 & .955 & \textbf{.993} \\ \hline
		\end{tabular}
	}
\end{table}

\begin{figure}[htbp]
	\centering
	\includegraphics[width= 0.8\textwidth]{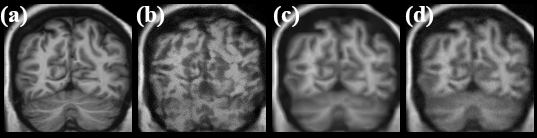}
	\caption{Selected Portion of MRI Reconstruction (20 \% Noisy samples): (a) 
		Original Image I6 (b) DAGAN 
		(50\% of training set) result (c) COROSA result (d) DL-MRI result}
	\label{fig:dagan_noisy_20per}
\end{figure}

In the next part, we repeated the first part with k-space samples corrupted by 
AWGN.  The noise variance was adjusted 
such that a specific  SNR is achieved with the zero-filled inverse transformed 
images.  These specific
SNR's  were chosen to be  0.5dB  less than that of noise-free  counter-parts
for the case of $10\%$  sampling, and to be  1.5dB  less  for the case of 
$20\%$  sampling.
The results are shown in Table \ref{table:4}, which demonstrate  that COROSA 
outperforms both DAGAN and DL-MRI 
in terms of both SNR and SSIM scores, except for two cases. In these two 
cases,  SNR of DAGAN is better.
It should be re-emphasized that, in 
terms of SSIM scores, COROSA is better than both DAGAN and DL-MRI
in all test cases. 
Figure \ref{fig:dagan_noisy_20per} shows the  reconstruction results 
corresponding to $20 \%$ sampling,  
which confirms the superiority of COROSA reconstruction.

\begin{table}[ht]
	\caption{Comparison of MRI reconstruction results with DAGAN and DL-MRI 
	(Noisy Measurements)}
	\label{table:4}
	\centering
	\scalebox{1}{
		\begin{tabular}{|c|c|c|c|c|c|c|c|c|c|c|c|}
			\hline
			\multirow{3}{*}{IMG} & \multirow{3}{*}{SR} & 
			\multicolumn{5}{c|}{SNR} & \multicolumn{5}{c|}{SSIM} \\ 
			\cline{3-12} 
			&  & \multirow{2}{*}{\begin{tabular}[c]{@{}c@{}}DL-\\ 
			MRI\end{tabular}} & \multicolumn{3}{c|}{DAGAN} & 
			\multirow{2}{*}{COR.} & 
			\multirow{2}{*}{\begin{tabular}[c]{@{}c@{}}DL-\\ MRI\end{tabular}} 
			& \multicolumn{3}{c|}{DAGAN} & \multirow{2}{*}{COR.} \\ \cline{4-6} 
			\cline{9-11}
			&  &  & 10\% & 50\% & 100\% &  &  & 10\% & 50\% & 100\% &  \\ \hline
			\multirow{2}{*}{I1} & .1 & 16.16 & 19.61 & 19.42 & 19.50 & 
			\textbf{19.64} & .349 & .738 & .837 & .820 & \textbf{.922} \\ 
			\cline{2-12} 
			& .2 & 19.91 & \textbf{24.47} & 24.67 & 25.21 & 22.72 & .475 & .845 
			& .890 & .888 & \textbf{.956} \\ \hline
			\multirow{2}{*}{I2} & .1 & 16.43 & 18.67 & 18.80 & 19.07 & 
			\textbf{19.40} & .363 & .718 & .819 & .806 & \textbf{.911} \\ 
			\cline{2-12} 
			& .2 & 19.80 & \textbf{24.22} & 24.06 & 24.89 & 22.60 & .481 & .847 
			& .883 & .890 & \textbf{.952} \\ \hline
			\multirow{2}{*}{I3} & .1 & 16.77 & 18.52 & 17.91 & 18.37 & 
			\textbf{18.60} & .393 & .717 & .801 & .793 & \textbf{.905} \\ 
			\cline{2-12} 
			& .2 & 19.48 & 22.13 & 20.92 & 21.61 & \textbf{22.16} & .452 & .795 
			& .784 & .769 & \textbf{.938} \\ \hline
			\multirow{2}{*}{I4} & .1 & 16.08 & 17.48 & 16.75 & 17.52 & 
			\textbf{18.19} & .398 & .702 & .783 & .777 & \textbf{.902} \\ 
			\cline{2-12} 
			& .2 & 19.38 & 20.69 & 19.39 & 19.68 & \textbf{21.56} & .448 & .746 
			& .715 & .698 & \textbf{.932} \\ \hline
			\multirow{2}{*}{I5} & .1 & 15.41 & 16.81 & 16.25 & 16.75 & 
			\textbf{17.70} & .394 & .696 & .769 & .766 & \textbf{.889} \\ 
			\cline{2-12} 
			& .2 & 19.36 & 20.90 & 19.32 & 19.42 & \textbf{21.20} & .457 & .757 
			& .706 & .696 & \textbf{.925} \\ \hline
			\multirow{2}{*}{I6} & .1 & 14.87 & 16.84 & 16.09 & 16.53 & 
			\textbf{17.34} & .391 & .686 & .757 & .749 & \textbf{.880} \\ 
			\cline{2-12} 
			& .2 & 19.52 & 20.54 & 19.28 & 19.26 & \textbf{21.17} & .473 & .762 
			& .729 & .708 & \textbf{.924} \\ \hline
		\end{tabular}
	}
\end{table}

With regards to the computation time, we ran all MATLAB algorithms on Core 
i7-3770 CPU 
with 8 GB RAM. We found that COROSA restoration for $256 \times 256$ 
image is completed in 185.7s. For the same task, TGV2, TV2, HS, COTV, COHS, 
DAGAN and DL-MRI take 24.2s, 2.3s, 2.9s, 2.8s, 3.6s, 0.2s and 507s 
respectively. Here, DAGAN restoration was performed using a trained network and 
the training itself requires approximately 4 hours for 10\% training set, 20 
hours for 50\% training set and 40 hours for the full training set of 15972 
images. 
Regarding overall computational complexity, our algorithm can be considered to 
be of 
$O(N^{2}\log(N))$ complexity for an $N \times N$ image, because of FFT 
operations that dominate over other filtering operations.

For detailed analysis, COROSA can be viewed mainly in two parts: 
the multiresolution loop involving alternative refinements of the required 
image and the adaptive weights 
iterations   (equations (\ref{eq:mrbeta})  and (\ref{eq:mrsi}))
and  the block coordinate descent iteration 
involving alternative refinements of the required image and the adaptive 
weights in the final resolution
(equations (\ref{eq:bcd1})  and (\ref{eq:bcd2})).  The first part is 
essentially similar to the second except
for the cost of expansion operation.  Ignoring the cost of expansion 
operation,  
one round of refinement of the image and the adaptive weight costs  
30+83$N_{i}$+44$N_{CG}N_{i}$ additions, 
35+78$N_{i}$+43$N_{CG}N_{i}$ multiplications,  $N_{i}$+1 square root operations 
and 2+6$N_{i}$+2$N_{CG}N_{i}$ FFT operations, where $N_i$ and $N_{CG}$ are the 
number of ADMM iterations and 
the number of CG iterations corresponding to the minimization of equation 
(\ref{eq:mrsi}), respectively.  
This complexity is multiplied by $N_f+K$ where $K$ is the number of 
multiresolution levels, and
$N_f$ is the number of block-coordinate descent iterations  (equations 
(\ref{eq:bcd1})  and (\ref{eq:bcd2})).    

\section{Conclusion}
\label{sec:conclusion}

We developed a novel form of regularization scheme that combines first- and 
second-order
derivatives in a manner that is adaptive to the image structure.   The 
adaptation is achieved
by the fact that the relative weight that combines first- and second-order 
derivatives is determined
by the same cost functional along with an additional regularization for 
preventing rapid variations
in the adaptive weights.  We used isotropic TV for the first-order term, and 
Hessian-Schatten norm
for the second order term.   We constructed an iterative method for minimizing 
the resulting non-convex 
and non-differentiable cost functional,  and we proved the convergence of the 
iterative method.
We demonstrated that, the proposed regularization method outperforms notable 
regularization methods
in the literature when the noise is high in the case of deblurring for  
microscopy. 
Further, in MRI reconstruction,  we demonstrated 
that the proposed regularization method outperforms 
existing   regularization methods,  and a recently proposed
Deep GAN method
when the noise and/or  under-sampling   are
high.  

\section*{Appendix}
\subsection*{Proof of Proposition 1}
Proof:  Solution $\bar{\beta}(\r) \in [0,1]^{N\times N}$ is given by
\begin{align*} 
	\bar{\beta}(\r) & = \underset{\beta}{\operatorname{argmin}} \;\;
	R_{sa}^{(j)}(s,\beta, p)  +  L(\beta, \tau) \\
	& = \underset{\beta}{\operatorname{argmin}} \;\; 
	\sum_{\bf r}\beta({\bf r}) \|({\bf d}_1*(E^{(j)}s))({\bf r}) \|_2  \\
	& + \sum_{\bf r} (1-\beta({\bf r}))\left\|{\pmb \zeta} 
	(({\bf H}_2*(E^{(j)}s))({\bf r}))\right\|_p \\
	& - \tau \sum_{\bf r}  \log\left(\beta(\r)(1-\beta(\r))\right).
\end{align*}
For the case  where $d(\r) = \|({\bf d}_1*(E^{(j)}s))({\bf r}) \|_2 
- \left\|{\pmb \zeta} (({\bf H}_2*(E^{(j)}s))({\bf r}))\right\|_p=0$,
equating derivative w.r.t.  $\beta$ to zero gives
$2\tau\bar{\beta(\r)} = \tau$
which gives the solution $\bar{\beta}(\r)=0.5$. 
Now we examine the case where $d(\r)$ is non-zero.
For this, we first define the following:
\begin{align}
	\nonumber
	v_{1}(\r)  &= \|({\bf d}_1*(E^{(j)}s))({\bf r}) \|_2, \\
	v_{2}(\r)  &= \left\|{\pmb \zeta} (({\bf H}_2*(E^{(j)}s))({\bf 
	r}))\right\|_p.
\end{align}
Next, we note that $\bar{\beta}(\r)$ is less than $0.5$  if $v_{1}(\r)  > 
v_{2}(\r)$
and greater than $0.5$  if $v_{1}(\r)  <  v_{2}(\r)$.  Hence,   the following
transformation  will reduce the range of minimization variable to $[0,0.5]$:
\begin{equation}
	\label{eq:shiftedbeta}
	\beta(\r)   = \frac{1}{2} - sign(d(\r))\beta_{r}(\r).
\end{equation}
Translating the optimization problem in terms of ${\beta}_{r}(\r)$
gives
\begin{align*}
	\bar{\beta}_{r}(\r) &= \underset{\beta}{\operatorname{argmin}} \;\;
	(0.5 + \beta_{r}(\r)) d_{l}(\r) + (0.5 - \beta_{r}(\r)) d_{h}(\r) 
	\ - \tau \log \left(	(0.5+ \beta_{r}(\r)) (0.5 - \beta_{r}(\r))\right) \\
	&= \underset{\beta}{\operatorname{argmin}} 
	-\beta_{r}(\r)(d_{h}-d_{l})(\r) - \tau \log \left( 0.25 - \beta_{r}^{2}(\r) 
	\right).
\end{align*} 
where
\begin{align*}
	d_{l}(\r)  = min(v_1(\r), v_2(\r)), \;\;  \mbox{and}\;\;
	d_{h}(\r)  = max(v_1(\r), v_2(\r)).
\end{align*}
Equating derivative w.r.t.  $\beta_{r}(\r)$  to zero  yields the 
quadratic equation
\begin{equation*}
	\beta_{r}^{2}(\r) + \frac{2\tau}{|d(\r)|} \beta_{r}(\r) - 0.25 = 0.
\end{equation*}
The roots are given by 
\begin{equation*}
	\beta_{r}^{+}(\r) = \frac{1}{2} \left( \sqrt{\zeta^2 + 1} - \zeta\right), \;
	\beta_{r}^{-}(\r) = \frac{1}{2} \left(  - \sqrt{\zeta^2 + 1} - \zeta\right).
\end{equation*}
where $\zeta = \frac{2\tau}{|d(\r)|}$.    From the form of the
expressions,  it is clear that $\beta_{r}^{+}(\r)$ is positive.  Also, as
$|d(\r)|$ ranges from $0$ to $\infty$,  $\beta_{r}^{+}(\r)$  
ranges from $0$ to $0.5$.  Hence,   
$\beta_{r}^{+}(\r) \in [0,0.5]^{N\times N}$, which is the required solution.
Substituting $\beta_{r}^{+}(\r)$ in the equation
\eqref{eq:shiftedbeta} gives the expression of the 
equation (\ref{eq:betaexp}). Since $\beta_{r}^{+}(\r) \in [0,0.5]$,  it is
clear that $\beta(\r)$ is guaranteed to be in the range $[0,1]$.   
In fact, it is  guaranteed to be in the interval $(0,1)$ because $L$ will be
infinity if $\beta(\r)\in \{0,1\}$.

\subsection*{Proof of Proposition 2}
The matrix in the equation \eqref{eq:ahatj} can be expressed as
\begin{equation}
	\label{eq:ahatj2}
	\begin{split}
		\hat{\mathbf{A}}_j  & = \mathbf{E}^{(j)T}\left(
		\mathbf{I} + \mathbf{D}_x^T{\bf D}_x + \mathbf{D}_y^T{\bf D}_y \right. 
		\left.  + \mathbf{D}_{xx}^T{\bf D}_{xx} + 
		\mathbf{D}_{yy}^T{\bf D}_{yy} + 
		2\mathbf{D}_{xy}^T{\bf D}_{xy} 
		\right)\mathbf{E}^{(j)}
	\end{split}
\end{equation}
Note that  $\mathbf{E}^{(j)}$ is the matrix equivalent of $j$-stage 
implementation
of two-fold upsampling;  this upsampling is realized as an expansion
by
a factor of two, which is inserting a zero after each pair of samples along both
axes,  and  then filtering by $u(\r) = 
\frac{1}{64}[1\;4\;6\;4\;1]^T[1\;4\;6\;4\;1]$. 
This operation is also equivalent to the single stage implementation involving
$2^j$ fold expansion,   and then filtering by  
$u_j(z_1,z_2)=\prod_{i=0}^{j-1}u(z^{2^i}_1,z^{2^i}_2)$  
where $u(z_1,z_2)$ is the $z$-transform of $u(\r)$ \cite{vaidfbtutorial}.
Let $u_j(\r)$  be the inverse
$z$-transform of $u_j(z_1,z_2)$.    From the structure of $\mathbf{E}^{(j)}$,
we infer that multiplication by $\mathbf{E}^{(j)T}$ is equivalent to 
convolution by
$u_j(-\r)$ followed by decimation by a factor of $2^j$ along both axes, which is
the operation of skipping $2^j-1$ samples for each block of $2^j$ samples.

Next,  the matrix ${\bf D}_x$   represents convolution by $d_x(\r)$ and  
${\bf D}_x^T$   represents convolution by $d_x(-\r)$. The other matrices within 
the 
square brackets are similarly interpreted.  

Hence the operation equivalent to
multiplication by $\hat{\mathbf{A}}_j$  is the following three stage operations
in sequence: (i)  $2^j$ fold expansion;  (ii) filtering by 
$B(\r)= 
u_j(\r)*u_j(-\r)*[1+d_x(\r)*d_x(-\r)+d_y(\r)*d_y(-\r)+d_{xx}(\r)*d_{xx}(-\r)
+d_{yy}(\r)*d_{yy}(-\r)+2d_{xy}(\r)*d_{xy}(-\r)]$; (iii) $2^j$ fold decimation.
This  three stage operation is equivalent to convolving by  $2^j$ fold decimated
version of $B(\r)$ \cite{vaidfbtutorial}.

\subsection*{Proof of Proposition 3}

Our proof will be based on Zangwill's global convergence theorem.  It states
three conditions to be satisfied by the   iterates to ensure convergence.
These conditions   translated    for our problem are the following:  (i)  the 
sequence 
$\{(s_{(k)},\beta_{(k)}\}_{i=1,2,\ldots}$ is a descent sequence,  i.e.,  the 
sequence
should satisfy  
$J_{sa}(s_{(k+1)},\beta_{(k+1)}, \tau, h, m) < 
J_{sa}(s_{(k)},\beta_{(k)}, \tau, h, m)$;  (ii)   the sequence of iterates 
should be
contained in a compact set;  (iii)  the mapping that generates the iterates
should be closed,  i.e.,  if $\mathcal{M}$ is mapping such that 
$(s_{(k+1)},\beta_{(k+1)}) = \mathcal{M}(s_{(k)},\beta_{(k)})$, then it should 
be
a closed mapping. 

To verify the second condition,  note that   
$\{(s_{(k)},\beta_{(k)}\}_{i=1,2,\ldots}$
is within the sub-level set  satisfying
$J_{sa}(s,\beta, \tau, h, m) \le J_{sa}(s_{(0)},\beta_{(0)}, \tau, h, m)$.
This is a bounded set because the function $J_{sa}(s,\beta, \tau, h, m)$ is
bounded below and has empty null space.  Note that a bounded set in
Euclidean  space is compact.

Note that $J_{sa}(s_{(k)},\beta_{(k+1)}, \tau, h, m) < 
J_{sa}(s_{(k)},\beta_{(k)}, \tau, h, m)$. This is because $\beta_{(k+1)}$ is 
computed
by exact of minimization of $J_{sa}(s_{(k)},\beta, \tau, h, m)$ with respect to 
$\beta$.   Next, $s_{(k+1)}$  is computed by iterative minimization of 
$J_{sa}(s,\beta_{(k+1)}, \tau, h, m)$ with respect to $s$ using ADMM.  
By assumption,
$J_{sa}(s_{(k+1)},\beta_{(k+1)}, \tau, h, m) < 
J_{sa}(s_{(k)},\beta_{(k+1)}, \tau, h, m)$. Hence we, have 
$J_{sa}(s_{(k+1)},\beta_{(k+1)}, \tau, h, m) < 
J_{sa}(s_{(k)},\beta_{(k)}, \tau, h, m)$. This verifies the first condition.

To verify the third condition, we will first verify that each cycle of ADMM
is a continuous mapping.  We first consider the mapping 
$(\mathbf{d}_f^{(k+1)}, \mathbf{d}_s^{(k+1)}, \mathbf{d}_0^{(k+1)})
= \mathcal{K}(\bar{\mathbf{s}}^{(k)}, \bar{\beta})$ represented by the equations
\eqref{eq:dfmin}, \eqref{eq:dsmin}, and \eqref{eq:d0min}.  Since these equations
represent exact single step minimizations of convex sub-functionals,  the 
mapping
$\mathcal{K}(\cdot)$ is continuous.  Next, consider the mapping 
$\bar{\mathbf{s}}^{(k+1)} = 
\mathcal{L}(\mathbf{d}_f^{(k+1)}, \mathbf{d}_s^{(k+1)}, \mathbf{d}_0^{(k+1)}, 
\bar{\beta})$
represented by the equation \eqref{eq:smin}.  Here, we ignore the presence of
$\mathbf{w}^{(k)}$ since it is generated by a simple affine transformation.
This minimization is implemented by
conjugate gradient iterations.    Conjugate gradient iteration with any
number of steps is equivalent to a minimization of convex quadratic function
within a subspace and it is also continuous.  
Hence the mapping, $\mathcal{M}_2(\cdot, \bar{\beta}) = 
\mathcal{L}(\mathcal{K}(\cdot, \bar{\beta}),\bar{\beta})$,  which represents
one cycle of ADMM,   is continuous.  
If the minimization given in equation \eqref{eq:bcd2}  is implemented with
$N_a$  cycles of ADMM  with initialization $s_{(k)}$,  we can represent this as 
$s_{(k+1)} = \mathcal{M}_2^{N_a}(s_{(k)}, \bar{\beta})$.  Now,  the minimization
in the equation \eqref{eq:bcd1}   is continuous operation because it is 
implemented
by exact minimization,  and function with respect to $\beta$ alone is convex.
The result, $\bar{\beta}$,  is a function of   $s_{(k)}$.  
We denote the minimization operation specified by the equation
\eqref{eq:bcd1},  by $(s_{(k)}, \bar{\beta}) =   \mathcal{M}_1(s_{(k)})$.
Hence one cycle of block coordinate descent can be represented
as $s_{(k+1)} = \mathcal{M}_2^{N_a}(\mathcal{M}_1(s_{(k)}))$. The above
mapping is also continuous since it is composition of continuous mappings.
Since continuity is a special case of closedness, we can say that
the mapping that generates the iterates is a closed mapping.

\subsection*{References for MRI Images}
\begin{flushleft}
	{\small
		MRI 1, https://www.xraygroup.com.au/index.php/our-services/mri \\
		MRI 2, https://www.healthcare.siemens.es/magnetic-resonance-imaging/options-and-upgrades/upgrades/magnetom-trio-upgrade/use \\
		MRI 3, https://radiopaedia.org/images/208569 \\
		MRI 4, https://www.usoccdocs.com/contract-service-mri/ \\
		MRI 5, https://www.researchgate.net/post/\url{What_do_you_think_diagnosis_of_this_pediatric_brain_MRI2} \\
		MRI 6, https://www.healthcare.siemens.co.uk/magnetic-resonance-imaging/0-35-to-1-5t-mri-scanner/magnetom-aera/use
	}	
\end{flushleft}

\bibliographystyle{unsrt}  
\bibliography{corosa}

\begin{thebibliography}{10}

\bibitem{CS_Baraniuk}
R.~G. Baraniuk, T.~Goldstein, A.~C. Sankaranarayanan, C.~Studer,
  A.~Veeraraghavan, and M.~B. Wakin.
\newblock Compressive video sensing: Algorithms, architectures, and
  applications.
\newblock {\em IEEE Signal Processing Magazine}, 34(1):52--66, Jan 2017.

\bibitem{Microscopy_Mvel}
Muthuvel Arigovindan, Jennifer~C. Fung, Daniel Elnatan, Vito Mennella,
  Yee-Hung~Mark Chan, Michael Pollard, Eric Branlund, John~W. Sedat, and
  David~A. Agard.
\newblock High-resolution restoration of {3D} structures from widefield images
  with extreme low signal-to-noise-ratio.
\newblock {\em Proceedings of the National Academy of Sciences},
  110(43):17344--17349, 2013.

\bibitem{astronomy_restoration}
Luxin Yan, Mingzhi Jin, Houzhang Fang, Hai Liu, and Tianxu Zhang.
\newblock Atmospheric-turbulence-degraded astronomical image restoration by
  minimizing second-order central moment.
\newblock {\em IEEE Geoscience and Remote Sensing Letters}, 9(4):672--676,
  2012.

\bibitem{Lustig_SparseMRI}
Michael Lustig, David Donoho, and John~M Pauly.
\newblock Sparse {MRI}: The application of compressed sensing for rapid {MR}
  imaging.
\newblock {\em Magnetic Resonance in Medicine: An Official Journal of the
  International Society for Magnetic Resonance in Medicine}, 58(6):1182--1195,
  2007.

\bibitem{Vogel_TV_98}
Curtis~R Vogel and Mary~E Oman.
\newblock Fast, robust total variation-based reconstruction of noisy, blurred
  images.
\newblock {\em IEEE transactions on image processing}, 7(6):813--824, 1998.

\bibitem{TV_App_BlindDecon}
T.~F. Chan and Chiu-Kwong Wong.
\newblock Total variation blind deconvolution.
\newblock {\em IEEE Transactions on Image Processing}, 7(3):370--375, March
  1998.

\bibitem{TV_App_Wavelet}
Sylvain Durand and Jacques Froment.
\newblock Reconstruction of wavelet coefficients using total variation
  minimization.
\newblock {\em SIAM Journal on Scientific computing}, 24(5):1754--1767, 2003.

\bibitem{TV_App_Wlinpaint}
Tony~F. Chan, Jianhong Shen, and Hao-Min Zhou.
\newblock Total variation wavelet inpainting.
\newblock {\em Journal of Mathematical Imaging and Vision}, 25(1):107--125, Jul
  2006.

\bibitem{k-svd}
Michal Aharon, Michael Elad, Alfred Bruckstein, et~al.
\newblock {K-SVD}: An algorithm for designing overcomplete dictionaries for
  sparse representation.
\newblock {\em IEEE Transactions on signal processing}, 54(11):4311, 2006.

\bibitem{MRI_Bresler}
Saiprasad Ravishankar and Yoram Bresler.
\newblock {MR} image reconstruction from highly undersampled k-space data by
  dictionary learning.
\newblock {\em IEEE transactions on medical imaging}, 30(5):1028, 2011.

\bibitem{Manifold_prior}
Jie Ni, Pavan Turaga, Vishal~M Patel, and Rama Chellappa.
\newblock Example-driven manifold priors for image deconvolution.
\newblock {\em IEEE Transactions on Image Processing}, 20(11):3086--3096, 2011.

\bibitem{DAGAN}
G.~{Yang}, S.~{Yu}, H.~{Dong}, G.~{Slabaugh}, P.~L. {Dragotti}, X.~{Ye},
  F.~{Liu}, S.~{Arridge}, J.~{Keegan}, Y.~{Guo}, and D.~{Firmin}.
\newblock {DAGAN}: Deep de-aliasing generative adversarial networks for fast
  compressed sensing {MRI} reconstruction.
\newblock {\em IEEE Transactions on Medical Imaging}, 37(6):1310--1321, June
  2018.

\bibitem{Deep_MRI_1}
Jiulong Liu, Tao Kuang, and Xiaoqun Zhang.
\newblock Image reconstruction by splitting deep learning regularization from
  iterative inversion.
\newblock In Alejandro~F. Frangi, Julia~A. Schnabel, Christos Davatzikos,
  Carlos Alberola-L{\'o}pez, and Gabor Fichtinger, editors, {\em Medical Image
  Computing and Computer Assisted Intervention -- MICCAI 2018}, pages 224--231,
  Cham, 2018. Springer International Publishing.

\bibitem{Deep_MRI_2}
Lei Xiang, Yong Chen, Weitang Chang, Yiqiang Zhan, Weili Lin, Qian Wang, and
  Dinggang Shen.
\newblock Ultra-fast {T2}-weighted {MR} reconstruction using complementary
  {T1}-weighted information.
\newblock In Alejandro~F. Frangi, Julia~A. Schnabel, Christos Davatzikos,
  Carlos Alberola-L{\'o}pez, and Gabor Fichtinger, editors, {\em Medical Image
  Computing and Computer Assisted Intervention -- MICCAI 2018}, pages 215--223,
  Cham, 2018. Springer International Publishing.

\bibitem{Deep_MRI_3}
Pengyue Zhang, Fusheng Wang, Wei Xu, and Yu~Li.
\newblock Multi-channel generative adversarial network for parallel magnetic
  resonance image reconstruction in k-space.
\newblock In Alejandro~F. Frangi, Julia~A. Schnabel, Christos Davatzikos,
  Carlos Alberola-L{\'o}pez, and Gabor Fichtinger, editors, {\em Medical Image
  Computing and Computer Assisted Intervention -- MICCAI 2018}, pages 180--188,
  Cham, 2018. Springer International Publishing.

\bibitem{DL_Caballero14}
Jose Caballero, Anthony~N Price, Daniel Rueckert, and Joseph~V Hajnal.
\newblock Dictionary learning and time sparsity for dynamic {MR} data
  reconstruction.
\newblock {\em IEEE transactions on medical imaging}, 33(4):979--994, 2014.

\bibitem{DL_mubarakhigher}
Minha Mubarak, Thomas~James Thomas, and Deepak Mishra.
\newblock Higher order dictionary learning for compressed sensing based dynamic
  {MRI} reconstruction.
\newblock {\em British Machine Vision Conference (BMVC)}, 2019.

\bibitem{tikh_1}
A.~N. Tikhonov and V.~Y. Arsenin.
\newblock Solution of ill-posed problems.
\newblock {\em V.H. Winston, Washington, DC}, 1977.

\bibitem{Rudin_tv1}
Leonid~I. Rudin, Stanley Osher, and Emad Fatemi.
\newblock Nonlinear total variation based noise removal algorithms.
\newblock {\em Physica D: Nonlinear Phenomena}, 60(1–4):259 -- 268, 1992.

\bibitem{TV_Edge}
David Strong and Tony Chan.
\newblock Edge-preserving and scale-dependent properties of total variation
  regularization.
\newblock {\em Inverse problems}, 19(6):S165, 2003.

\bibitem{TV_Ring_Stair}
Wolfgang Ring.
\newblock Structural properties of solutions to total variation regularization
  problems.
\newblock {\em ESAIM: Mathematical Modelling and Numerical Analysis},
  34(4):799--810, 2000.

\bibitem{TV_Stair_Jalalzai}
Khalid Jalalzai.
\newblock Some remarks on the staircasing phenomenon in total variation-based
  image denoising.
\newblock {\em Journal of Mathematical Imaging and Vision}, 54(2):256--268, Feb
  2016.

\bibitem{TV_Chambolle_97}
Antonin Chambolle and Pierre-Louis Lions.
\newblock Image recovery via total variation minimization and related problems.
\newblock {\em Numerische Mathematik}, 76(2):167--188, 1997.

\bibitem{Scherzer_tv2_98}
O.~Scherzer.
\newblock Denoising with higher order derivatives of bounded variation and an
  application to parameter estimation.
\newblock {\em Computing}, 60(1):1--27, 1998.

\bibitem{TV_Chan_2000}
Tony Chan, Antonio Marquina, and Pep Mulet.
\newblock High-order total variation-based image restoration.
\newblock {\em SIAM Journal on Scientific Computing}, 22(2):503--516, 2000.

\bibitem{HS_13}
S.~Lefkimmiatis, J.~P. Ward, and M.~Unser.
\newblock Hessian schatten-norm regularization for linear inverse problems.
\newblock {\em IEEE Transactions on Image Processing}, 22(5):1873--1888, 2013.

\bibitem{HS_12}
Stamatios Lefkimmiatis, Aur{\'e}lien Bourquard, and Michael Unser.
\newblock Hessian-based norm regularization for image restoration with
  biomedical applications.
\newblock {\em IEEE Transactions on Image Processing}, 21(3):983--995, 2012.

\bibitem{vogel_tv1}
CR~Vogel.
\newblock Total variation regularization for ill-posed problems.
\newblock {\em Department of Mathematical Sciences Technical Report}, 1993.

\bibitem{TV_Dual_Chambolle}
Antonin Chambolle.
\newblock An algorithm for total variation minimization and applications.
\newblock {\em Journal of Mathematical imaging and vision}, 20(1-2):89--97,
  2004.

\bibitem{tvadmm}
Yilun Wang, Junfeng Yang, Wotao Yin, and Yin Zhang.
\newblock A new alternating minimization algorithm for total variation image
  reconstruction.
\newblock {\em SIAM Journal on Imaging Sciences}, 1(3):248--272, 2008.

\bibitem{TV_Aujol}
Jean-Fran{\c{c}}ois Aujol.
\newblock Some first-order algorithms for total variation based image
  restoration.
\newblock {\em Journal of Mathematical Imaging and Vision}, 34(3):307--327,
  2009.

\bibitem{TV_CombinedLysaker}
Marius Lysaker and Xue-Cheng Tai.
\newblock Iterative image restoration combining total variation minimization
  and a second-order functional.
\newblock {\em International Journal of Computer Vision}, 66(1):5--18, 2006.

\bibitem{Combined_order_TV}
K.~Papafitsoros and C.~B. Sch{\"o}nlieb.
\newblock A combined first and second order variational approach for image
  reconstruction.
\newblock {\em J. Math. Imaging Vision}, 48(2):308--338, 2014.

\bibitem{TGV}
Kristian Bredies, Karl Kunisch, and Thomas Pock.
\newblock Total generalized variation.
\newblock {\em SIAM Journal on Imaging Sciences}, 3(3):492--526, 2010.

\bibitem{sam-tv}
S.~Viswanath, S.~de~Beco, M.~Dahan, and M.~Arigovindan.
\newblock Multi-resolution based spatially adaptive multi-order total variation
  for image restoration.
\newblock In {\em 2018 IEEE 15th International Symposium on Biomedical Imaging
  (ISBI 2018)}, pages 497--500, April 2018.

\bibitem{TGV2}
Florian Knoll, Kristian Bredies, Thomas Pock, and Rudolf Stollberger.
\newblock Second order total generalized variation ({TGV}) for {MRI}.
\newblock {\em Magnetic Resonance in Medicine}, 65(2):480--491, 2011.

\bibitem{eckstein1992douglas}
Jonathan Eckstein and Dimitri~P. Bertsekas.
\newblock On the {D}ouglas-{R}achford splitting method and the proximal point
  algorithm for maximal monotone operators.
\newblock {\em Mathematical Programming}, 55(1):293--318, Apr 1992.

\bibitem{eckstein2017approximate}
Jonathan Eckstein and Wang Yao.
\newblock Approximate {ADMM} algorithms derived from {L}agrangian splitting.
\newblock {\em Computational Optimization and Applications}, 68(2):363--405,
  2017.

\bibitem{unsersigpromag}
Michael Unser, Akram Aldroubi, and Murray Eden.
\newblock B-spline signal processing. i. theory.
\newblock {\em IEEE transactions on signal processing}, 41(2):821--833, 1993.

\bibitem{Bertsekas_NLP}
Dimitri~P Bertsekas.
\newblock {\em Nonlinear programming}.
\newblock Athena scientific Belmont, 1999.

\bibitem{TV_ADMM_Afonso}
M.~V. {Afonso}, J.~M. {Bioucas-Dias}, and M.~A.~T. {Figueiredo}.
\newblock Fast image recovery using variable splitting and constrained
  optimization.
\newblock {\em IEEE Transactions on Image Processing}, 19(9):2345--2356, Sep.
  2010.

\bibitem{TV_ADMM_Steidl}
G.~Steidl and T.~Teuber.
\newblock Removing multiplicative noise by {D}ouglas-{R}achford splitting
  methods.
\newblock {\em Journal of Mathematical Imaging and Vision}, 36(2):168--184, Feb
  2010.

\bibitem{TV_ADMM_Manu}
M.~{Ghulyani} and M.~{Arigovindan}.
\newblock Fast total variation based image restoration under mixed
  {P}oisson-{G}aussian noise model.
\newblock In {\em 2018 IEEE 15th International Symposium on Biomedical Imaging
  (ISBI 2018)}, pages 1264--1267, April 2018.

\bibitem{proximal_parikh}
Neal Parikh, Stephen Boyd, et~al.
\newblock Proximal algorithms.
\newblock {\em Foundations and Trends{\textregistered} in Optimization},
  1(3):127--239, 2014.

\bibitem{bertero_inverse}
Mario Bertero and Patrizia Boccacci.
\newblock {\em Introduction to inverse problems in imaging}.
\newblock CRC press, 1998.

\bibitem{ssim}
Zhou Wang, Alan~C Bovik, Hamid~R Sheikh, and Eero~P Simoncelli.
\newblock Image quality assessment: from error visibility to structural
  similarity.
\newblock {\em IEEE transactions on image processing}, 13(4):600--612, 2004.

\bibitem{MRI_Chauffert}
Nicolas Chauffert, Philippe Ciuciu, Jonas Kahn, and Pierre Weiss.
\newblock Variable density sampling with continuous trajectories.
\newblock {\em SIAM Journal on Imaging Sciences}, 7(4):1962--1992, 2014.

\bibitem{vaidfbtutorial}
Parishwad~P Vaidyanathan.
\newblock Multirate digital filters, filter banks, polyphase networks, and
  applications: a tutorial.
\newblock {\em Proceedings of the IEEE}, 78(1):56--93, 1990.

\end{thebibliography}

\end{document}